\documentclass[aps,prb,preprint,superscriptaddress,nofootinbib,longbibliography,amsmath,amssymb]{revtex4-2}
\usepackage{graphicx,booktabs,array}
\newcolumntype{P}[1]{>{\raggedright\arraybackslash}p{#1}}
\makeatletter
\g@addto@macro\@mkpream@relax{\let\@endpbox\relax\let\do@row@strut\relax\let\ar@align@mcell\relax}
\makeatother
\usepackage{placeins,flafter}
\usepackage{url}
\PassOptionsToPackage{hyphens}{url}
\usepackage[hidelinks]{hyperref}
\usepackage{enumitem}
\setlist{nosep,leftmargin=1.6em}

\newcommand{\G}{\mathrm{G8}}
\newcommand{\Oref}{\mathrm{cuBLAS}}

\newcommand{\meanpm}[2]{#1\,\mathord{\pm}\,#2}
\graphicspath{{figures/}{./}}

\begin{document}
\sloppy

\title{GPU acceleration of the numerical-atomic-orbital DFT code OpenMX with GEMMul8: performance and numerical agreement on consumer and HPC GPUs}

\author{Hiroyuki Kawai}
\email{h.river1234@gmail.com}
\affiliation{Department of Physics, Niigata University, Niigata 950-2181, Japan}
\affiliation{X-Ability Co., Ltd, Bunkyo-ku, Tokyo 113-0033, Japan}
\author{Taisuke Ozaki}
\affiliation{Institute for Solid State Physics, The University of Tokyo, Kashiwa, Chiba 277-8581, Japan}
\author{Yoshiaki \=Ono}
\affiliation{Department of Physics, Niigata University, Niigata 950-2181, Japan}

\date{November 9, 2026}

\begin{abstract}
We integrate GEMMul8, a library that emulates FP64 matrix multiplication on INT8 Tensor Cores, into the GPU version of the numerical-atomic-orbital first-principles code OpenMX 4.0. The GEMMul8 configuration is compared against the cuBLAS FP64 configuration within each of two environments, a consumer GeForce RTX 5080 and an HPC-oriented H100 PCIe, for the four computational paths formed by collinear (CL) or noncollinear (NCL) spin and band or $\Gamma$-point-cluster solvers. In fixed-work measurements that completed 25 SCF iterations, the diagonalization section of a 384-atom CL band calculation on the RTX 5080 is accelerated by a factor of 2.1 and the total run time by 1.8; a 512-atom CL cluster calculation, which uses only real matrix products in contrast to the CL band calculation, still gains a factor of 1.5 in the diagonalization section. On the H100, GEMMul8 provides little additional acceleration of the diagonalization section. In an accuracy test on a 216-atom Si system in which one atom was displaced by 0.05~\AA, all 24 configurations were converged using an SCF convergence criterion of $10^{-8}$; every configuration converged at the same iteration within each path, and the differences between GEMMul8 and cuBLAS are at most $2\times10^{-12}$~Ha in the total energy and $7\times10^{-12}$~Ha/Bohr in force components.
\end{abstract}

\maketitle
\newpage

\section{Introduction}

Density functional theory (DFT)~\cite{hk,ks} is the standard first-principles method for the electronic structure of solids and molecules~\cite{burke2012,jones2015,jain2016}. Plane-wave implementations~\cite{payne1992,kresse1996,qe2009} are the most widespread, while localized-basis methods using Gaussian~\cite{frisch1984,weigend2005} or Slater-type~\cite{tevelde2001} functions are also common. Localized-basis methods based on numerical atomic orbitals (NAOs) exploit compact basis sets and locality in real space~\cite{sankey1989,siesta2002,blum2009,openmx}. In computational paths that rely on dense diagonalization, however, solving the generalized eigenvalue problem requires roughly $O(n^3)$ operations in the number of basis functions $n$, and it dominates the run time for large systems~\cite{sanchez1997,motamarri2014}. Graphics processing units (GPUs) offer high parallel throughput and wide memory bandwidth~\cite{owens2008,mittal2015}, and GPU acceleration of localized-basis codes has been reported for Gaussian-basis quantum-chemistry programs~\cite{yasuda2008a,yasuda2008b,wilkinson2011,williamsyoung2020} as well as for FHI-aims~\cite{huhn2020}, SIESTA~\cite{siesta-gpu}, ABACUS~\cite{abacus}, and OpenMX~\cite{parq2014,kawai2025}. How the matrix construction, the eigenvalue solution, and the density update are distributed over the GPU therefore determines the performance of the whole program~\cite{huhn2020,kawai2025}.

In our previous work~\cite{kawai2025} we introduced GPU acceleration based on the cuBLAS/cuSOLVER libraries~\cite{cublas,cusolver} into the collinear (CL) and noncollinear (NCL) band calculations of OpenMX. In OpenMX 4.0 GPU v2.0, the subject of the present study, the Hamiltonian construction, the real-space grid operations, the density-matrix evaluation, and the forces have been moved to the GPU in addition to the dense-matrix operations, and the downstream processing that consumes the eigenvectors obtained on a single GPU has been restructured. Progress in GPU coverage alone, however, does not remove the constraints that originate inside the eigensolver, such as the bandwidth-bound Householder reduction~\cite{haidar2011,yamazaki2014}. Moreover, the ratio between native FP64 (double-precision floating-point) throughput and low-precision matrix throughput differs from one GPU to another, so that a straightforward port of the same FP64 algorithm is not always effective~\cite{ozaki2,woo2026}. On consumer GPUs in particular, FP64 throughput is customarily capped at a small fraction of the FP32 (single-precision floating-point) throughput; for the GeForce RTX 5080 used here the theoretical peaks are 0.88~TFLOPS (FP64), 56.3~TFLOPS (FP32), and 450~TOPS (dense INT8, 8-bit integer operations), i.e.\ an $\mathrm{FP64}:\mathrm{FP32}:\mathrm{INT8}$ balance of $1:64:512$~\cite{uchino-complex}. As long as first-principles calculations of several hundred atoms depend on the FP64 units, this imbalance limits their speed on such GPUs. INT8 Tensor Core throughput, on the other hand, has grown from one GPU generation to the next; if it can be redirected to high-precision computation, such calculations could become practical on consumer GPUs costing about one to two thousand US dollars~\cite{ozaki2,uchino-fp8}.

Matrix-multiplication emulation based on the Ozaki scheme has been studied as a way of using low-precision matrix units for high-precision computation~\cite{ozaki2012,mukunoki2019,mukunoki2020,ootomo2024,uchino2025}. Since this scheme is central to the present work, we outline it here. The original scheme (Ozaki scheme I)~\cite{ozaki2012} is based on error-free transformations. In Tensor-Core emulation, the input matrices are approximated by sums of components with reduced significands,
\begin{equation}
 A\simeq\sum_{i=1}^{s_A}A^{(i)},\qquad B\simeq\sum_{j=1}^{s_B}B^{(j)},
 \label{eq:ozaki1-split}
\end{equation}
with appropriate power-of-two scaling so that the retained partial products can be evaluated without rounding error within the precision and range of the accumulator (for example FP16 inputs with FP32 accumulation, or INT8 inputs with INT32 accumulation). For equal split counts $s_A=s_B=s$, an FP64 approximation is formed from selected partial products, $AB\simeq\sum_{i+j\le s+1}A^{(i)}B^{(j)}$, with the final summation performed in FP64. Its accuracy depends on the number of splits, the omitted terms, and rounding in the final summation~\cite{mukunoki2020,ootomo2024}. Because the number of partial products grows quadratically with $s$, the gain of scheme I is limited when close to full FP64 accuracy is required.

Ozaki scheme II~\cite{ozaki2} removes this quadratic growth. The inputs are scaled row- or column-wise and rounded to integer matrices $A_I$ and $B_I$, and their products are evaluated modulo a set of pairwise coprime integers $p_1,\ldots,p_L$ (the moduli) by INT8 matrix units,
\begin{equation}
 R_\ell=\bigl(A_I\bmod p_\ell\bigr)\bigl(B_I\bmod p_\ell\bigr)\bmod p_\ell,
 \qquad \ell=1,\ldots,L .
 \label{eq:ozaki2-residue}
\end{equation}
By the Chinese remainder theorem, an integer whose magnitude is smaller than $M/2$ with $M=\prod_{\ell=1}^{L}p_\ell$ is uniquely determined by its residues, so that the exact integer product $A_IB_I$ can be reconstructed element-wise from $\{R_\ell\}$ and rescaled to floating point. The number of modular matrix products is therefore linear in $L$, and $L$ controls the accuracy: with about 15 moduli the emulated product can attain FP64-level accuracy for suitably scaled inputs, whereas the number of moduli actually required depends on the exponent distribution of the inputs, and returning output in FP64 must be distinguished from guaranteeing results identical to a particular FP64 implementation for arbitrary inputs~\cite{uchino-error}. GEMMul8 provides an implementation of scheme II with a cuBLAS-like interface, and its performance and power efficiency on INT8 matrix engines have been reported~\cite{gemmul8,uchino-sc25ws}; an extension to complex matrix products has also been reported~\cite{uchino-complex}, which makes the approach applicable not only to real symmetric problems but to electronic-structure calculations with wave-vector dependence and spinor representations. Improved scaling for its fast mode~\cite{kawakami2026}, an FP8 (8-bit floating-point) variant~\cite{uchino-fp8}, an extension to FP4 Tensor Cores~\cite{hayashi2026}, and variants with guaranteed accuracy~\cite{guaranteed} have also been reported.

Reduced precision requirements in electronic-structure calculations have also been examined for density-matrix evaluation~\cite{dawson2024} and for mixed-precision iterative eigensolvers~\cite{woo2026}. Mixed-precision methods choose the precision per operation and combine it with high-precision corrections where necessary; they do not necessarily sacrifice the accuracy of the final physical quantities~\cite{abdelfattah2021,higham2022}. In density fitting for quantum chemistry, adaptive-precision INT8 arithmetic has been used while preserving converged energies~\cite{huang2026}. Liu \textit{et al.}\ applied precision emulation to MuST/LSMS through automatic BLAS offloading~\cite{liu2026}. In contrast to these, the present work targets a scheme in which the explicit DGEMM/ZGEMM calls in the OpenMX source are switched while the FP64 input/output is retained. This scheme allows the replaced matrix shapes, the call counts, the workspace management, and the fallbacks to be evaluated in correspondence with the computational path.

The aim of this work is not only to show the stand-alone performance of GEMMul8 but also to clarify how much of its gain is reflected in the total run time of NAO-based DFT. For the four computational paths we contrast the consumer RTX 5080 with the HPC-oriented H100, and discuss the cuBLAS/GEMMul8 comparison within each environment, the verification of numerical agreement, a performance model based on stand-alone GEMM measurements, and the constraints imposed by memory and concurrent execution. The CPU-versus-GPU comparison is presented separately as an evaluation of the current GPU implementation as a whole and is not confused with the contribution of GEMMul8 alone.

\section{Computational paths and the integration of GEMMul8}

\subsection{Standardization of the generalized eigenvalue problem}

Omitting the wave-vector and spin indices, the Kohn--Sham equation in an NAO basis reads
\begin{equation}
 H C=S C\varepsilon ,
\end{equation}
where $H$ is the Hamiltonian matrix, $C$ the matrix of expansion coefficients, $S$ the overlap matrix, and $\varepsilon$ the diagonal matrix of eigenvalues. When $S$ is positive definite, writing $S=U\Lambda U^\dagger$ and
\begin{equation}
 X=U\Lambda^{-1/2},\qquad X^\dagger S X=I,\qquad
 \widetilde H=X^\dagger H X
 \label{eq:standardization}
\end{equation}
gives $C=XY$ from the solution of the standard eigenvalue problem $\widetilde H Y=Y\varepsilon$. This is one representation of the standardization based on L\"owdin orthogonalization~\cite{lowdin1950}; it is related to the representation using the symmetric orthogonalization $S^{-1/2}=U\Lambda^{-1/2}U^\dagger$ by a unitary transformation within the orthogonal basis. The dense matrix products required for the standardization and the back-transformations are the main parts replaced in this work. The amplification of errors caused by an ill-conditioned overlap matrix is a separate issue from the stand-alone error of a matrix product.

In NCL calculations the matrix dimension in the spinor basis is $n_2=2n$, and $X_2=\operatorname{diag}(X,X)$ is used for the spin-independent overlap matrix. In this paper NCL denotes the computational path that uses spinors and is distinguished from whether spin--orbit coupling (SOC) is included. The real symmetric or Hermitian eigenvalue problem after standardization is solved with the existing cuSOLVER path~\cite{cusolver}. Our implementation does not modify the operations inside cuSOLVER and switches only the explicit GEMMs outside it to GEMMul8; this work is therefore not a reduced-precision treatment of the eigensolver itself.

\subsection{Matrix products in the four computational paths}

Hereafter, band/CL, band/NCL, cluster/CL, and cluster/NCL are abbreviated as bcol, bnc, ccol, and cnc, followed by the number of atoms; bcol384, for example, denotes a 384-atom CL band calculation. ``Cluster'' refers to the \texttt{cluster} eigenvalue solver of OpenMX. The Si cluster paths in this paper use $\Gamma$-point calculations of periodic Si supercells, not isolated Si clusters constructed in vacuum. The undisplaced Si of the performance tests and the displaced Si of the accuracy tests are treated as distinct atomic structures (Section~\ref{sec:test-structures}).

The GEMM composition of each path is given in Table~\ref{tab:gemm}. Band calculations use ZGEMM because complex matrices appear at general $k$ points, whereas CL $\Gamma$-point calculations use DGEMM. In the NCL cluster, the real and imaginary parts of the three spin blocks that are independent by Hermitian symmetry are handled separately; two-sided transformations of six real matrices use two DGEMMs each, so the standardization amounts to 12 DGEMMs. The back-transformation uses the $n_2$-dimensional complex eigenvectors, so DGEMM and ZGEMM coexist.

\begin{table}[tbp]\centering\small
\caption{Main GEMMs replaced, where $m$ is the number of computed states (\texttt{MaxN} in the code), $n$ the number of NAO basis functions, and $n_2=2n$. The rightmost column gives the call counts for 25 SCF iterations in the performance tests on undisplaced Si, with eight computed $k$ points for band paths; the 64-atom systems that terminated early are excluded.}
\label{tab:gemm}
\begin{tabular}{@{}lP{0.60\linewidth}c@{}}\toprule
Path & Main matrix products per dense solve & Calls per run\\\midrule
bcol & three $n$-dimensional ZGEMMs & 600 Z\\
bnc & six $n$-dimensional ZGEMMs and one $n_2$-dimensional back-transformation & 1400 Z\\
ccol & two $n$-dimensional DGEMMs and one back-transformation of shape $(m,n,n)$ & 75 D\\
cnc & twelve $n$-dimensional DGEMMs and one ZGEMM back-transformation of shape $(m,n_2,n_2)$ & 300 D + 25 Z\\\bottomrule
\end{tabular}
\end{table}

Since the number of states $m$ may differ between the first and the subsequent SCF iterations, the performance model distinguishes non-square back-transformations from square products. For cnc216, for instance, the back-transformation of the first iteration and that of the subsequent iterations with $m=1987$ and $n_2=5616$ are counted as different shapes. Counting a real matrix product conventionally as $2mnk$ operations and a complex one as $8mnk$, the twelve DGEMMs of cnc amount to $24n^3$ and the ZGEMM back-transformation to $8mn_2^2$; their ratio depends on $m/n_2$ and is not a fixed fraction.

In band calculations the MPI ranks are divided into groups per $k$ point, and the dense matrices of each group are processed by one GPU-owning rank; when eight computed $k$ points are processed by a single GPU, several owning ranks therefore share the same GPU. In the cluster calculations with \texttt{gpusolver} used in this work, the dense-matrix part is processed on a single GPU and the remaining operations are distributed with MPI. The reference path \texttt{gpusolver2} distributes the dense-matrix processing of the cluster itself using ELPA GPU and COSMA~\cite{elpa,yu2021,cosma}.

\subsection{FP64 emulation by integer matrix products}

The emulation follows Ozaki scheme II as outlined in the Introduction: the inputs are scaled and rounded to integer matrices $A_I$ and $B_I$, their residue products $R_\ell$ are evaluated by INT8 matrix units for pairwise coprime moduli $p_1,\ldots,p_L$ according to Eq.~\eqref{eq:ozaki2-residue}, and the integer product $A_IB_I$ is reconstructed by the Chinese remainder theorem~\cite{ozaki2}. With $M=\prod_\ell p_\ell$, the integer values are recovered uniquely provided each element of the integer product lies in the corresponding signed recovery range, e.g.\ $|(A_IB_I)_{ij}|<M/2$. This integer-recovery property must, however, be distinguished from the errors associated with the quantization and rescaling of the original floating-point input; the required number of moduli depends on the exponent distribution of the inputs and on the required accuracy~\cite{uchino-error}.

Our measurements used the INT8 backend of GEMMul8 v3.2.0 (commit \texttt{833e5761})~\cite{gemmul8,uchino-sc25ws}, with \texttt{num\_moduli}=15 for both real and complex products and with fast mode disabled; this accuracy setting was fixed for all system sizes. We do not, on the other hand, claim a universal error bound for arbitrary DFT inputs from the mere fact that 15 moduli were used; the numerical verification is presented in Section~\ref{sec:accuracy}.

\subsection{Switching, workspace, and recording of the execution path}

The OpenMX side provides cuBLAS-compatible real and complex GEMM wrappers, and \path{scf.gemmul8.enable} switches between cuBLAS FP64 and GEMMul8 within the same binary; complex products use the complex interface of GEMMul8. In our measurements the additional effect of GEMMul8 was evaluated by switching only this while leaving the other GPU processing unchanged.

The workspace is managed per $(\text{device},\text{stream})$ within each MPI process. If the requested size exceeds a prescribed fraction of the total VRAM, if the free memory after allocation would fall below a prescribed reserve, or if the allocation fails, the corresponding GEMM falls back to cuBLAS. The fraction can be changed with \path{OPENMX_GEMMUL8_MAX_WORKSPACE_PERCENT}; it is a \emph{limit applied to each request}, not a limit on the sum of the workspaces of all processes sharing the GPU. In the practical-convergence re-measurement of the displaced Si216 bnc case, both the RTX 5080 and the H100 used a 50\% limit and a 1536~MiB reserve of free memory after allocation; the original tight-threshold bnc accuracy test on the RTX 5080 used the same setting. For the other series the run conditions are identified by the effective values recorded in the individual manifests. This memory-protection fallback is not a mechanism that detects numerical errors or guarantees accuracy.

The execution paths of the main performance measurements, the tight-threshold accuracy tests, and the practical-convergence re-measurements were identified from the per-run \texttt{manifest.json} and the corresponding calculation logs. The manifest records the build identifier, the MPI/OpenMP layout, the number of SCF iterations and the convergence verdict, the numbers of dense GPU/CPU solves, the number of successful GEMMul8 calls, the numbers of real and complex fallbacks with their reasons, and the observed memory values, because requesting the GPU in the input and actually executing the target operation on the GPU or with GEMMul8 are not the same thing.

\section{Benchmarks and evaluation method}

\subsection{Computational environments and input conditions}

The environments of the main measurements are listed in Table~\ref{tab:env}. Both used a single GPU, flat MPI, and one OpenMP thread per rank. The CUDA Multi-Process Service (MPS) was enabled in the regular GPU measurements, and only the MPS on/off comparison was run as a separate series~\cite{mps}. CPU reference calculations used the CPU path (ELPA2) of the same executable. Since the H100 and RTX 5080 environments also differ in CPU, memory capacity, and rank count, the absolute times of the two environments are not treated as a comparison of the GPU products alone.

\begin{table}[tbp]\centering\small
\caption{Environments of the main measurements. The GB figures of product names are distinguished from the observed memory values quoted in binary prefixes (GiB) in the text. The Pegasus configuration was cross-checked against the published specification of the facility~\cite{pegasus}. The code row gives the commit from which all executables were built; the tag \texttt{v2.0\_thesis} was later moved to a commit with identical source files (see Data availability).}
\label{tab:env}
\begin{tabular}{@{}l|P{0.36\linewidth}|P{0.36\linewidth}@{}}\hline
Item & Pegasus (CCS, Univ.\ of Tsukuba) & Workstation\\\hline
GPU & H100 PCIe 80 GB, one per node & GeForce RTX 5080 16 GB, one\\
CPU & Xeon Platinum 8468, 48 cores & Core i9-10980XE, 18 cores\\
Host memory & DDR5 128 GiB per node & 256 GiB\\
MPI ranks & 48 ranks per node in the main measurements & 16 or 18 ranks depending on the series\\\hline
Software & \multicolumn{2}{l@{}}{NVHPC 26.5, CUDA 13.2, GEMMul8 v3.2.0, OpenMPI}\\
Code & \multicolumn{2}{l@{}}{built from \texttt{v2.0\_thesis} (\texttt{db72cd4})}\\\hline
\end{tabular}
\end{table}

On the RTX 5080, the performance measurements of undisplaced Si bcol216, ccol216, and cnc216 were run with 18 ranks (all physical cores), and the other main points including bnc216 with 16 ranks. In preliminary measurements comparing the two on the same systems, the diagonalization times agreed within 0.4~s (0.3\%), because the dense GPU work does not depend on the host rank count, whereas the total run times differed by 3--4\% for the cluster paths and by at most 2\% for the band paths. Table~\ref{tab:5080} gives the rank count of each RTX 5080 point in Fig.~\ref{fig:speedup}, and points with different rank counts are not treated as one strong-scaling series.

The performance tests used diamond-structure Si supercells without deliberate atomic displacements (hereafter undisplaced Si), with the GGA-PBE exchange--correlation functional~\cite{pbe}, the \texttt{Si7.0-s2p2d1} basis, and the \texttt{Si\_PBE19} pseudopotential of DFT\_DATA19. The number of basis functions for this basis is $n=13N$. The real-space grid cutoff was 200~Ry and the electronic temperature 300~K. Band calculations used a $2\times2\times2$ $k$-point mesh, i.e.\ the eight computed $k$ points of this implementation, while the cluster paths used the $\Gamma$ point only; total energies of different paths are therefore not compared as if they resulted from the same $k$-point integration.

In the Si CL series \texttt{scf.SpinPolarization} was \texttt{off}, and in the NCL series it was \texttt{NCL}. SOC was not included in any of the Si performance tests, band or cluster, including the NCL series. The CL series is thus an evaluation of the spin-unpolarized case, and the NCL series is a performance test of the computational path that passes through the spinor matrix representation without SOC. These Si results are not taken as a verification of the performance of SOC calculations or of the convergence of general magnetic materials. SOC was enabled in the supplementary NCL tests on Cr$_2$ and GaAs, whose conditions differ from those of Si (Section~\ref{sec:accuracy}).

The cell shapes and lattice vectors are kept in the input files; for instance, a $4\times4\times4$ supercell of the primitive cell was used for 128 atoms and a $3\times3\times3$ supercell of the conventional cell for 216 atoms. Not only the number of atoms but also the cell shape can affect the local operations and the memory usage.

In the final main measurements on the RTX 5080, bcol216, ccol216, and cnc216 were run with 18 ranks and the other main points with 16 ranks; this corresponds to the final aggregation \path{s5prod_thesis_result_v2.txt} of the measurement archive. At each comparison point the rank counts of cuBLAS and GEMMul8 were identical, but the series was not measured with a single rank count over all sizes. The RTX 5080 accuracy-test series was run entirely with 16 ranks and is distinguished from the performance tests.

\subsection{Test structures and the purposes of the evaluations}
\label{sec:test-structures}

Undisplaced Si denotes inputs in which each supercell was built from the reference coordinates of the diamond structure without local atomic displacements for accuracy verification. Displaced Si216, on the other hand, is a $3\times3\times3$ supercell of the conventional cell with edge 16.293~\AA{} in which only atom~1 was moved by 0.05~\AA{} along $x$: in fractional coordinates the $x$ component of atom~1 was changed from 0.04166667 to 0.04473547 while the coordinates of the other atoms and the cell were kept. This displacement provides a structure with non-zero forces, so that the backend differences of force components as well as energies could be evaluated. Neither name implies that the structure was relaxed by geometry optimization.

\begin{table}[tbp]\centering\small
\caption{Structures and purposes of the tests used in this paper. The performance tests on undisplaced Si and the accuracy tests on displaced Si216 are distinguished. SOC is disabled in all Si tests.}
\label{tab:test-structures}
\begin{tabular}{@{}P{.24\linewidth}P{.22\linewidth}P{.48\linewidth}@{}}\toprule
Test & Structure or input & Object of evaluation\\\midrule
Main performance, MPS, parallelization, memory & undisplaced Si supercells & Run times etc.\ were compared under the condition that 25 SCF iterations were completed; early-terminating cases were separated.\\
Tight-threshold accuracy test & displaced Si216 (atom~1 moved by 0.05~\AA{} along $+x$) & 24 runs (4 paths, 2 environments, 3 configurations). Stress test with threshold $10^{-13}$ and a 60-iteration cap; final physical quantities and SCF histories were compared.\\
Practical-convergence re-measurement & the same displaced Si216 & The same 24 configurations were run once each with threshold $10^{-8}$ and a 200-iteration cap. The actual terminating iteration, the energies and forces at convergence, and the differences from the original stress test were evaluated.\\
Matrix-level auxiliary verification & undisplaced Si64 & Matrices and eigenpairs saved at the third SCF iteration of a CL $\Gamma$-point calculation were compared. Separate from the final-output accuracy tests on displaced Si216.\\
Stand-alone GEMM test & fixed-seed uniform random matrices & Times and numerical differences of matrix products with the call shapes of the DFT runs. Not an accuracy test of an atomic structure.\\
Small auxiliary tests & MnO, Cr$_2$, GaAs & MnO in AFM CL, Cr$_2$ and GaAs in NCL+SOC. Consistency of the existing computational paths without GEMMul8 calls was checked.\\\bottomrule
\end{tabular}
\end{table}

Even for the same path name and number of atoms, the SCF histories of undisplaced and displaced Si can differ. The terminating iterations of 20 or 23 in the practical-convergence re-measurement refer to the accuracy-test inputs of displaced Si216; they are not convergence counts common to the performance tests on undisplaced Si or to arbitrary Si216 inputs.

\subsection{Fixed work versus early termination}
\label{sec:fixedwork}

The performance inputs specified \texttt{scf.maxIter}=25 and \texttt{scf.criterion}=$10^{-15}$~Ha~\cite{openmx-manual} with the intention of executing 25 SCF iterations. \texttt{scf.maxIter} is, however, an upper bound and not a setting that forces the iteration count. In the regular fixed-basis band and cluster calculations of the \texttt{v2.0\_thesis} source of OpenMX 4.0 GPU used for the measurements, the convergence flag is raised from the second SCF iteration on when the band-energy difference \texttt{dUele} between consecutive iterations falls below \texttt{scf.criterion} and, at the same time, the most recent entry \texttt{NormRD[0]} of the residual-norm history array (index 0 denotes the latest mixing step, not an initial value) is, at the check, below ten times that value~\cite{openmxgpu-source}; termination is therefore not judged by the energy difference alone. The correspondence between this check and the printed output for the density mixing used here is given in Section~\ref{sec:practical-convergence}.

According to the run records, the cuBLAS configurations of bnc64 and cnc64 on the RTX 5080 terminated after 20 and 22 iterations, respectively, whereas their GEMMul8 configurations ran to 25. These two cases are listed separately in Table~\ref{tab:early} and are excluded from the fixed-25-iteration speedups, from the speedup-versus-size figure, and from the evaluation of the Amdahl-type model. Since dividing the time by the iteration count does not separate the costs of initialization, the first diagonalization, and post-processing, no conversion to fixed work was attempted either. Completion of 25 iterations was confirmed for all other main points. These performance values are run times for a fixed number of iterations and do not directly compare the time to a converged solution.

\begin{table}[tbp]\centering\small
\caption{Undisplaced Si64 on the RTX 5080 (16 ranks, five runs each), separated from the fixed-work comparison. Times are means $\pm$ sample standard deviations. Because the amounts of work differ, no ranking by speedup is drawn.}
\label{tab:early}
\begin{tabular}{lcccc}\toprule
 & \multicolumn{2}{c}{SCF iterations} & \multicolumn{2}{c}{Total run time [s]}\\
System & cuBLAS & GEMMul8 & cuBLAS & GEMMul8\\\midrule
bnc64 & 20 & 25 & $\meanpm{86.09}{0.40}$ & $\meanpm{89.21}{0.30}$\\
cnc64 & 22 & 25 & $\meanpm{58.55}{0.20}$ & $\meanpm{63.50}{0.43}$\\
\bottomrule\end{tabular}
\end{table}

\subsection{Timing, statistics, and comparison metrics}

Run times were taken from \texttt{Max\_Time} in the timing table at the end of the output file, i.e.\ the maximum over MPI ranks. In this paper \texttt{Total Computational Time} is called the total run time and \texttt{Diagonalization} the diagonalization section; the latter is not the cuSOLVER call alone but includes the standardization GEMMs, the eigenvalue solution, the back-transformation, and the other processing inside the routine. Since \texttt{MD.maxIter}=1 was specified, the total run time also includes initialization and one force evaluation, which differs from the steady-state step time of a long molecular-dynamics run.

In the main measurements, five runs were taken per condition for the GPU configurations and three per condition for the CPU reference and the MPS-off configurations. We report arithmetic means and sample standard deviations, and the speedups in tables and figures are ratios of the arithmetic means of the two configurations,
\begin{equation}
 S_\G^{q}=\frac{\overline T_\Oref^{q}}{\overline T_\G^{q}},
 \qquad q\in\{\mathrm{total},\mathrm{diag}\},
 \label{eq:speedup}
\end{equation}
where a value larger than one means that GEMMul8 is faster. CPU-versus-GPU speedups are obtained by dividing the mean time of the CPU reference by the mean time of the GPU configuration in the same environment with the same rank count. The sample standard deviation is the run-to-run spread, not the standard error of the mean or a confidence interval. The time reduction is $1-1/S_\G^{\mathrm{total}}$ and is distinguished from the speedup.

Four groups in the main H100 series had a coefficient of variation of the total run time above 2\%, the largest being 4.23\% for the GEMMul8 configuration of~bcol300; all five runs were retained for these groups as well, and the spread is reported alongside. Preliminary tests and main measurements are separate series, and the statistics of this section are computed from all retained runs of the main measurements. The warm-up convention for the stand-alone GEMM measurements is described separately in Section~\ref{sec:micro}.

The supplementary multi-node series used five or seven runs. The measurement archive records the convention of excluding allocations that contained the known slow nodes \texttt{bnode013} and \texttt{bnode033} and re-measuring; the excluded raw data are preserved, and the numbers in the text correspond to the set retained under this convention. These results are therefore not regarded as an estimate of the operational distribution of the whole machine including anomalous or slow nodes.

\subsection{Conditions of the numerical verification}

For the displaced Si216 system defined in Section~\ref{sec:test-structures}, the four paths were compared on the same structure with non-zero forces. As in the performance tests on undisplaced Si, SOC was not included, including the NCL paths. The numerical verification consists of a comparison at practical convergence and a tight-threshold stress test that continues the iteration beyond practical convergence.

In the practical-convergence re-measurement, \texttt{scf.criterion}=$10^{-8}$~Ha and \texttt{scf.maxIter}=200 were used, and the four paths bcol, bnc, ccol, and cnc with the three configurations CPU, cuBLAS, and GEMMul8 were run once each in both the H100 and RTX 5080 environments, 24 runs in total. The H100 used 48 MPI ranks and the RTX 5080 16 MPI ranks, both with \texttt{-nt 1}; MPS was enabled for the GPU configurations. Within each environment the same \texttt{v2.0\_thesis} binary as in the original accuracy tests was used, and the atomic coordinates, cell, basis, pseudopotentials, temperature, cutoff, and density-mixing settings were kept. \texttt{MD.Opt.criterion} was $10^{-5}$~Ha/Bohr with a geometry-optimization cap of one step, so this is not a comparison of relaxed structures. Setting the SCF cap to 200, well above the predicted stopping iteration, separates termination by the cap from termination by convergence.

Prior to this, a stress test with \texttt{scf.criterion}=$10^{-13}$~Ha and a 60-iteration cap was carried out on the same 24 configurations. Its conditions and results are summarized in Appendix~\ref{app:stress}; from here on the practical-convergence re-measurement is the primary subject.

In this paper the pair of conditions on the energy difference and the density residual corresponding to \texttt{scf.criterion}=$10^{-8}$ is defined as the practical convergence criterion. This is a definition for the purpose of evaluation in the present tests, not an error guarantee common to arbitrary physical quantities. The re-measurement uses the actual SCF terminating iterations and the physical quantities at convergence, and the a posteriori first-reach evaluation from the original $10^{-13}$ histories was cross-checked separately.

The total energy is \texttt{Utot}, and the forces are all atoms and components of the \texttt{coordinates.forces} block. The between-backend comparison quantities at the same stopping threshold are defined as
\begin{equation}
 \Delta E_{A,B}=|E_A-E_B|,\qquad
 \Delta F_{\max,A,B}=\max_{I,\alpha}|F_{I\alpha,A}-F_{I\alpha,B}| ,
 \label{eq:accuracy}
\end{equation}
where the force comparison is not the difference between the magnitudes of the maximum forces. The energy differences of the re-measurement were obtained by subtracting the printed \texttt{Utot} values recorded in the aggregation as decimal numbers, so that no rounding from re-conversion to binary floating point is introduced; the maximum force differences use aggregated values comparing the corresponding atom and component. A zero means agreement within the printed precision, and no claim is made about differences below the printed digits.

The effect of the stopping threshold was evaluated by comparing, for the same environment, structure, path, and backend, the output at convergence with threshold $10^{-8}$ against the output at the end of the $10^{-13}$ stress test (iterations 47--60). This is referred to below as the between-threshold difference. Unlike the between-backend difference, it does not give the error with respect to the exact solution or the difference from a fully converged solution, because the $10^{-13}$ runs include cases that stopped at the cap.

\subsection{Terminating iterations of the practical-convergence re-measurement}
\label{sec:practical-convergence}

The primary observable of the re-measurement is the SCF terminating iteration of the runs that actually specified $10^{-8}$. It was read from the calculation logs, the aggregation records, and the convergence verdict in the corresponding manifests, and compared with the a posteriori values from the original histories. The bundled aggregation also contains a simplified auxiliary indicator that checks the energy difference alone except at the second iteration, but the definition of convergence in this paper is the pair of conditions of Eq.~\eqref{eq:scf-stop} and is not replaced by that indicator.

Table~\ref{tab:practical-convergence} lists the actual terminating iterations of all 24 configurations in the re-measurement. The first-reach iterations of the original stress test are cross-check values, not substitutes for the measured ones. As in the corresponding stopping test of OpenMX, satisfying the conditions in several consecutive iterations is not required. The physical quantities at practical convergence were evaluated from the final output of the re-measurement, while the terminating iterations and final quantities of the stress test were kept as separate results. Each configuration was re-measured once, and this is distinguished from statistical speedup evaluations based on repeated timing.

The details of the convergence test and the procedure used to obtain the first-reach iteration corresponding to $10^{-8}$ from the original stress-test histories are given in Appendix~\ref{app:stress}.

\section{Results}

\subsection{Effect of the current GPU implementation as a whole}

\begin{figure}[tbp]\centering
\includegraphics[width=.94\linewidth]{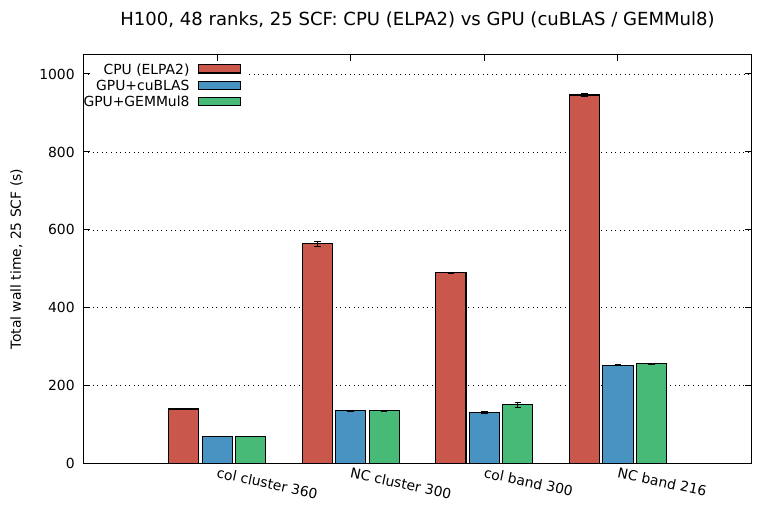}
\caption{Total run times of undisplaced Si on the same H100 node with 48 ranks. Bars are arithmetic means and error bars sample standard deviations (three CPU runs, five GPU runs each). Completion of 25 SCF iterations was confirmed for all conditions. The CPU-versus-GPU comparison and the cuBLAS-versus-GEMMul8 comparison on the GPU are distinct.}
\label{fig:layer1}
\end{figure}

On the same H100 node with 48 ranks, the total-run-time speedups of GPU (cuBLAS) over CPU (ELPA2) are 3.73 for bcol300, 3.75 for bnc216, 1.70 for ccol216, and 4.16 for cnc300 (Fig.~\ref{fig:layer1}); the range over the seven main conditions is 1.70--4.16. These values represent the effect of the current GPU implementation as a whole, not that of GEMMul8. The differences between solvers include the share of the diagonalization in the total and the performance differences between real and complex matrix processing on the CPU and on the GPU. The times of all seven conditions and the additional effect of GEMMul8 are given in Appendix~\ref{app:h100}.

\subsection{Additional effect of GEMMul8 on the RTX 5080}
\label{sec:rtx}

\begin{figure}[tbp]\centering
\includegraphics[width=\linewidth]{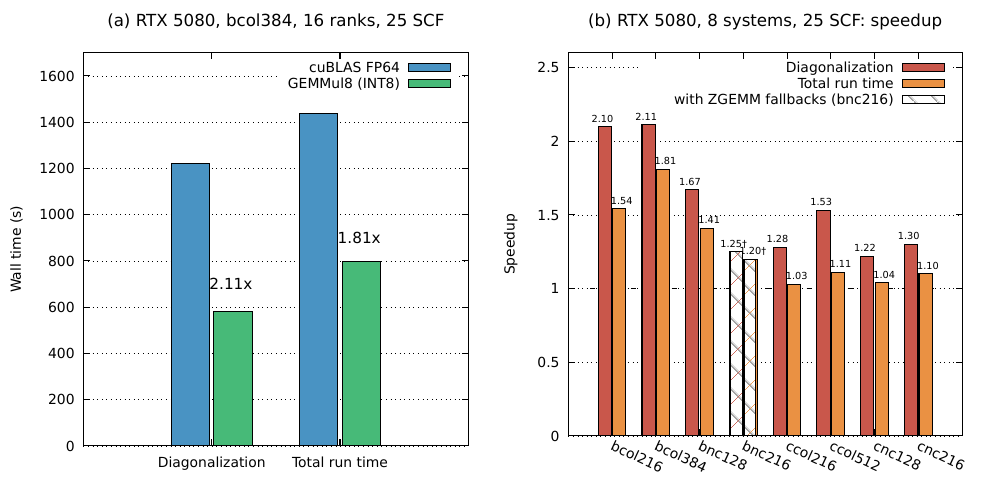}
\caption{Additional effect of GEMMul8 on the RTX 5080 at a fixed 25 SCF iterations (undisplaced Si, no SOC). (a) bcol384: wall time of the diagonalization section and of the whole run for cuBLAS FP64 and GEMMul8 (means of five runs); the labels give the speedups 2.11 and 1.81. (b) Diagonalization and total speedups $T_{\rm cuBLAS}/T_{\rm GEMMul8}$ for all eight fixed-work systems; the dashed line marks a speedup of one. The hatched bars and the dagger ($\dagger$) mark bnc216, whose GEMMul8 run includes 193 ZGEMM fallbacks to cuBLAS, so that its speedups are those of a mixed path (Table~\ref{tab:5080}).}
\label{fig:headline}
\end{figure}

Table~\ref{tab:5080} gives the RTX 5080 results that satisfy the fixed-25-iteration condition. The largest gain is obtained for bcol384, where the diagonalization time decreases from $1222.03$~s to $579.92$~s and the total run time from $1436.85$~s to $795.33$~s; the speedups are 2.11 and 1.81, and the reduction of the total run time is 44.65\%. For bcol216 the diagonalization section is also accelerated by 2.10, but because the non-diagonalization share is larger, the total speedup stays at 1.54. Figure~\ref{fig:headline} summarizes these gains: panel (a) shows the wall times of bcol384 for the two backends, and panel (b) the diagonalization and total speedups of all eight fixed-work systems.

\begin{table}[tbp]\centering\small
\caption{Comparison on the RTX 5080 at a fixed 25 SCF iterations. All systems are undisplaced Si without SOC, including the NCL paths. Five runs per configuration; times are means $\pm$ sample standard deviations, and speedups are ratios of means. $\dagger$: includes 193 ZGEMM fallbacks. The standard deviations of the total run times are listed in the re-aggregated CSV.}
\label{tab:5080}
\setlength{\tabcolsep}{3.7pt}
\begin{tabular}{lccccc}\toprule
System & ranks & \multicolumn{2}{c}{Diagonalization time [s]} & $S_\G^{\rm diag}$ & $S_\G^{\rm total}$\\
 & & cuBLAS & GEMMul8 & & \\\midrule
bcol216 & 18 & $\meanpm{235.46}{0.07}$ & $\meanpm{112.36}{0.13}$ & 2.10 & 1.54\\
bcol384 & 16 & $\meanpm{1222.03}{1.73}$ & $\meanpm{579.92}{0.37}$ & 2.11 & 1.81\\
bnc128 & 16 & $\meanpm{275.61}{0.97}$ & $\meanpm{164.63}{0.21}$ & 1.67 & 1.41\\
bnc216$^{\dagger}$ & 16 & $\meanpm{1262.98}{1.06}$ & $\meanpm{1009.21}{0.86}$ & 1.25 & 1.20\\
ccol216 & 18 & $\meanpm{13.98}{0.04}$ & $\meanpm{10.89}{0.03}$ & 1.28 & 1.03\\
ccol512 & 16 & $\meanpm{110.83}{0.04}$ & $\meanpm{72.32}{0.06}$ & 1.53 & 1.11\\
cnc128 & 16 & $\meanpm{34.14}{0.04}$ & $\meanpm{27.95}{0.03}$ & 1.22 & 1.04\\
cnc216 & 18 & $\meanpm{132.32}{0.03}$ & $\meanpm{101.80}{0.09}$ & 1.30 & 1.10\\
\bottomrule\end{tabular}
\end{table}

Even for ccol512, where only DGEMMs are replaced, the diagonalization speedup is 1.53, while the total speedup is 1.11. For cnc216, twelve DGEMMs for six real two-sided transformations and one complex back-transformation are replaced, giving 1.30 for the diagonalization section and 1.10 overall. These results show that emulating real matrix products alone already yields a gain, and that a large stand-alone GEMM gain does not translate directly into the total-run-time speedup. The times of the 64-atom systems are kept in Table~\ref{tab:early}, but because of the different iteration counts no ranking or crossover for small systems at fixed work is asserted from them.

For bnc216, fallbacks occurred under the prescribed memory-protection conditions. Of the 1400 ZGEMM requests per run, 1207 were executed by GEMMul8 and 193 fell back to cuBLAS, with the same counts in all five runs and with the reserve policy protecting the free memory after allocation as the reason in every case. By call site, 193 of the 200 back-transformations of size $5616^3$ returned to cuBLAS. Even this mixed path gave 1.25 for the diagonalization section and 1.20 overall, but the result is not one in which GEMMul8 executed all target products. In this paper bnc128 is used as the fallback-free representative of the NCL band.

\begin{figure}[tbp]\centering
\includegraphics[width=.94\linewidth]{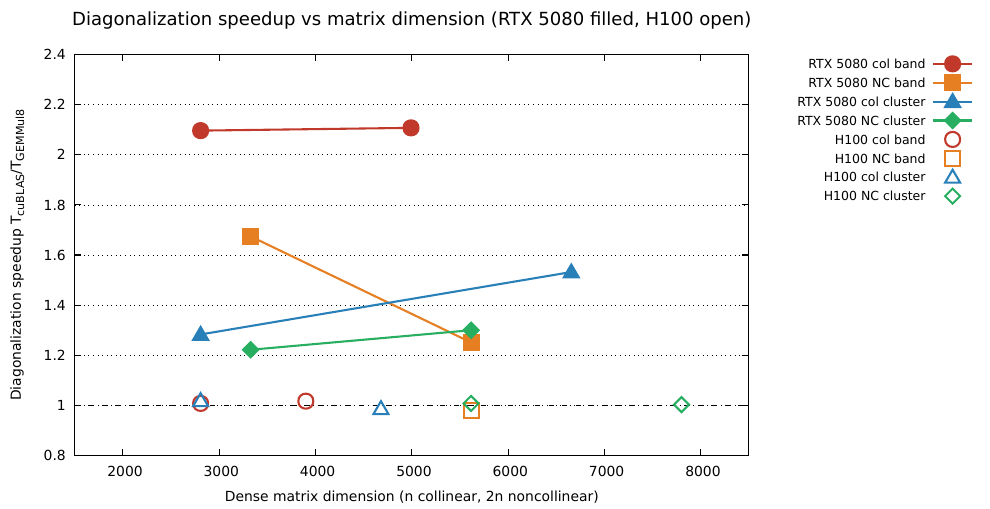}
\caption{Diagonalization speedup versus the dimension of the dense eigenvalue problem for undisplaced Si without SOC. Filled symbols: RTX 5080; open symbols: H100. All values are ratios of means, and values above one mean that GEMMul8 is faster. The early-terminating 64-atom cases are excluded. The RTX 5080 points use 16 or 18 ranks depending on the point and do not form a strong-scaling curve for one layout. The RTX 5080 point of bnc216 includes fallbacks.}
\label{fig:speedup}
\end{figure}

\subsection{Limited gain on the H100 and the effect on other processing}
\label{sec:h100}

In the main H100 series the diagonalization speedup is 0.978--1.019, with no large additional gain like that on the RTX 5080. ``Limited'' here means that no large acceleration was observed; it does not claim that the differences are statistically zero for all conditions. In the supplementary cnc384 series with 36 ranks and the pre-freeze binary, the mean diagonalization times are 95.61~s with cuBLAS and 93.63~s with GEMMul8, a speedup of about 1.02; this series is, however, not joined to the main series to define a universal matrix-dimension threshold (Appendix~\ref{app:provenance}).

\begin{figure}[tbp]\centering
\includegraphics[width=.87\linewidth]{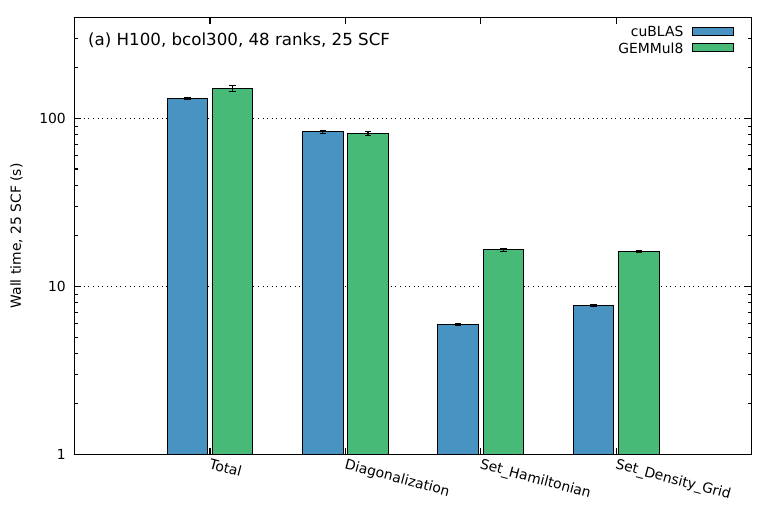}\\[2mm]
\includegraphics[width=.87\linewidth]{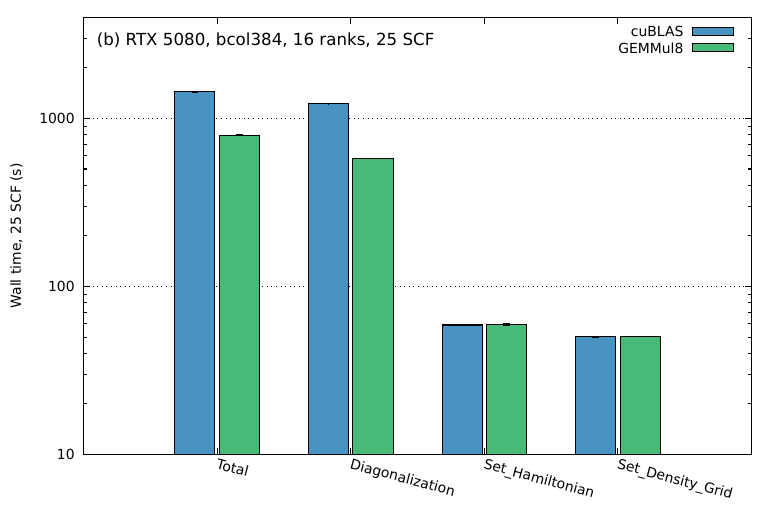}
\caption{Total run time and main processing sections of undisplaced Si (five runs each, mean $\pm$ sample standard deviation, logarithmic scale). For bcol300 on the H100 the increase outside the diagonalization cancels the gain. Because the rank attaining \texttt{Max\_Time} may differ between sections, the independently measured sections are shown side by side rather than stacked as an exact additive decomposition.}
\label{fig:phases}
\end{figure}

For bcol300 on the H100, although the diagonalization section becomes slightly shorter, from $83.04$~s to $81.65$~s, the total run time increases by 14.97\% from $131.37$~s to $151.03$~s: \texttt{Set\_Hamiltonian} increases from $5.95$~s to $16.56$~s and \texttt{Set\_Density\_Grid} from $7.71$~s to $16.15$~s (Fig.~\ref{fig:phases}). At the same time the maximum observed VRAM over the five runs rises from 56.77~GiB to 68.89~GiB, and eviction of resident data and a reduced number of concurrent executions were recorded. These observations are consistent with the interpretation that workspace allocation interacts with the memory use of other processing on a GPU shared by several $k$-point-owning ranks. Since this is not a causal experiment that isolates the contribution of each mechanism, the whole increase is not attributed to any particular one.

\subsection{Stand-alone GEMMs and an Amdahl-type model}
\label{sec:micro}

The stand-alone GEMM tests measured the actual call shapes and transposition flags as well as a series of square matrices for the size dependence: 46 conditions on the RTX 5080 and 59 on the H100 were measured and saved. The 105 conditions in total do not mean 105 distinct application shapes. On the RTX 5080 the 46 conditions consist of the 28 call shapes of the eight systems listed in Table~\ref{tab:amdahl} and 18 square matrices of the size sweep; the four call shapes of cnc128, namely the 1664-dimensional DGEMMs with their two transposition flags and the ZGEMM back-transformations of shapes $(1177,3328,3328)$ and $3328^3$, were measured on 1 November 2026, after the other 42 conditions of 27 August 2026, with the same program and procedure on the same GPU, against a GEMMul8 library of the same commit (\texttt{833e5761}) built from a fresh checkout of commit \texttt{db72cd4b}, then the target of the \texttt{v2.0\_thesis} tag. For the 1664-dimensional non-transposed DGEMM, which is common to the size sweep of the first date and the cnc128 set, the stand-alone times of the two dates differ by about 9\% for cuBLAS and 7\% for GEMMul8; this day-to-day spread is part of the uncertainty of the model below. The model is evaluated only with stand-alone times measured for the exact shapes and transposition flags of each run. The inputs were fixed-seed uniform random matrices with $\alpha=1,\beta=0$. After allocating buffers and workspace and warming up both backends, the two were measured alternately seven times each, and \emph{the median} was adopted for the stand-alone tests; this median plays a different role from the arithmetic mean of the repeated whole-SCF measurements.

In the square, non-transposed size sweep, the cuBLAS/GEMMul8 ratio on the RTX 5080 is 6.62--18.57 for DGEMM and 10.34--25.40 for ZGEMM, whereas on the H100 it is 0.21--1.09 for DGEMM and 0.25--1.52 for ZGEMM; over all measurements including non-square shapes, the largest H100 ZGEMM ratio is 1.55. The ratio also depends on the shape and the transposition flags, not on the matrix dimension alone. The normalized difference from cuBLAS obtained in the stand-alone tests is at most $1.4\times10^{-14}$ over all 105 conditions. This indicator is the maximum absolute difference of real components divided by the maximum absolute value of the real components of the reference output, with the real and imaginary parts of complex numbers counted as separate components; it differs from the maximum component-wise relative error over the non-zero elements.

To estimate the propagation to the application, let $j$ index the replaced shape and transposition flag, $c_j$ its total number of calls, $g_j$ the number of successful GEMMul8 calls, and $t_{j,\Oref}$ and $t_{j,\G}$ the stand-alone times, and form
\begin{align}
 A&=\sum_j c_jt_{j,\Oref},\\
 B&=\sum_j\{g_jt_{j,\G}+(c_j-g_j)t_{j,\Oref}\} .
 \label{eq:mixmodel}
\end{align}
Without fallbacks, $g_j=c_j$. For the diagonalization section we put
\begin{equation}
 \widehat f=\frac{A}{\overline T_\Oref^{\rm diag}},\qquad
 s_{\rm eff}=\frac{A}{B},\qquad
 S_{\rm pred}=\frac{1}{(1-\widehat f)+\widehat f/s_{\rm eff}} .
 \label{eq:amdahl}
\end{equation}
Here $\widehat f$ is a \emph{work-based estimate} obtained by summing stand-alone times, not a GEMM time fraction measured directly on the critical path of an application involving several ranks and asynchronous execution. The model rests on the approximation that the non-GEMM time is unchanged and that the stand-alone times per shape can be transferred into the run. The first-iteration and subsequent back-transformation shapes were treated separately, and no coefficients were fitted to the speedups.

\begin{table}[tbp]\centering\small
\caption{Amdahl-type model for the RTX 5080 performance tests on undisplaced Si. Measured values are ratios of the mean times of the main measurements. The prediction error is $100(S_{\rm pred}/S_{\rm meas}-1)$; bnc216 includes the measured fallback mix. The stand-alone times of cnc128 were measured on 1 November 2026, those of the other systems on 27 August 2026 (Section~\ref{sec:micro}).}
\label{tab:amdahl}
\begin{tabular}{lccccc}\toprule
System & $\widehat f$ & $s_{\rm eff}$ & $S_{\rm pred}$ & $S_{\rm meas}$ & Error [\%]\\\midrule
bcol216 & 0.549 & 18.81 & 2.083 & 2.096 & -0.61\\
bcol384 & 0.585 & 23.99 & 2.275 & 2.107 & +7.95\\
bnc128 & 0.465 & 14.27 & 1.763 & 1.674 & +5.28\\
bnc216 & 0.475 & 1.75 & 1.255 & 1.251 & +0.26\\
ccol216 & 0.243 & 12.33 & 1.288 & 1.283 & +0.34\\
ccol512 & 0.365 & 18.25 & 1.527 & 1.532 & -0.36\\
cnc128 & 0.222 & 11.91 & 1.255 & 1.222 & +2.69\\
cnc216 & 0.255 & 15.86 & 1.314 & 1.300 & +1.07\\
\bottomrule\end{tabular}
\end{table}

As shown in Table~\ref{tab:amdahl}, the predicted and measured diagonalization speedups agree within 1.5\% for five of the eight conditions. For bnc216 the prediction including the 193 fallbacks is 1.255 against a measured 1.251. Setting the fallbacks to zero while keeping the same RTX 5080 stand-alone and non-GEMM times gives a prediction of about 1.83, but this is a counterfactual estimate, not a measurement under changed memory conditions, and it cannot be read across to a prediction for the H100.

For bcol384, bnc128, and cnc128 the model overestimates the speedup by 7.95\%, 5.28\%, and 2.69\%, respectively. We do not call this a measured value of the integration overhead itself. The time-normalized residual
\begin{equation}
 \delta=\frac{\overline T_\G^{\rm diag}}{\overline T_\Oref^{\rm diag}}
       -\left(1-\widehat f+\frac{\widehat f}{s_{\rm eff}}\right)
\end{equation}
is 0.0349 for bcol384, 0.0300 for bnc128, and 0.0215 for cnc128. The residual may contain synchronization, concurrent execution, shared caches and bandwidth, and differences between the stand-alone and in-run conditions; the prediction error of the speedup and the actual fraction of added time are therefore not treated as the same number.

\begin{figure}[tbp]\centering
\includegraphics[width=.82\linewidth]{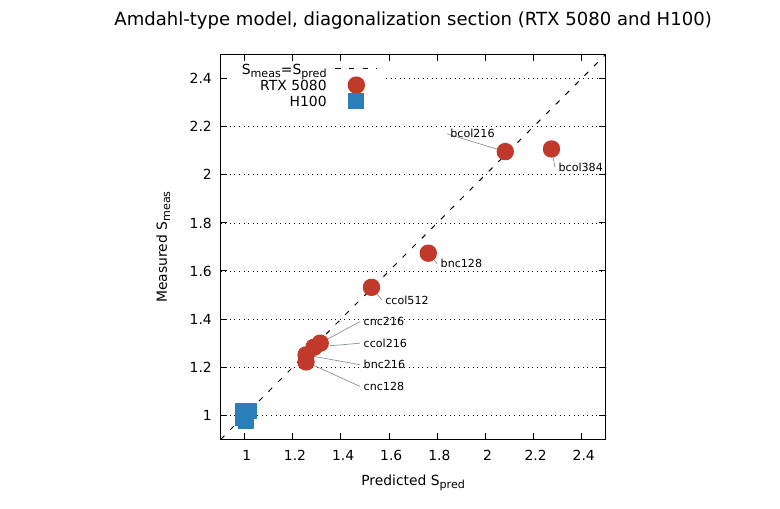}
\caption{Amdahl-type predictions versus measurements for the diagonalization section of undisplaced Si. The dashed line marks agreement. On the H100 both predictions and measurements cluster near a speedup of one. The model uses the work estimate based on stand-alone GEMM times.}
\label{fig:amdahl}
\end{figure}

On the H100, $\widehat f=0.0085$--$0.119$, smaller than the $0.222$--$0.585$ of the RTX 5080; the contrast between the two GPUs thus arises not only from the difference in $s_{\rm eff}$ but also from the different relative weight of GEMM work within the reference cuBLAS run. Furthermore, for bcol300 on the H100 the prediction for the diagonalization section is close to the measurement, whereas the prediction for the total run time does not capture the increase in the other processing. This result shows the need to distinguish the three levels: stand-alone GEMM, diagonalization section, and total run time.

\subsection{Behaviour with MPS and on multiple nodes}

\begin{figure}[tbp]\centering
\includegraphics[width=.94\linewidth]{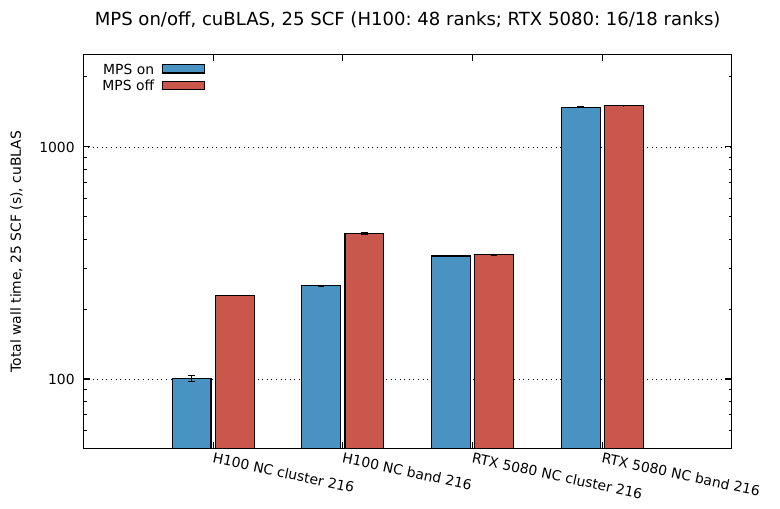}
\caption{MPS-off/on comparison of undisplaced Si at the same rank count within each system (H100: 48; RTX 5080: 16 for bnc216 and 18 for cnc216; cuBLAS, 25 SCF iterations). Times are means $\pm$ sample standard deviations, three MPS-off runs and five MPS-on runs; the vertical axis is logarithmic. Because the CPUs and rank counts differ between the GPUs, the effect of MPS is compared within each environment.}
\label{fig:mps}
\end{figure}

The effect of MPS is shown in Fig.~\ref{fig:mps}. The ratio of the total run time of the cuBLAS configuration without and with MPS is 1.68 for bnc216 and 2.28 for cnc216 on the H100 with 48 ranks, but 1.014 for bnc216 with 16 ranks and 1.015 for cnc216 with 18 ranks on the RTX 5080. This difference is consistent with a dependence of the MPS effect on the number of ranks sharing the GPU and on the processing structure, although the two environments also differ in hardware. MPS contributes substantially in the H100 layout studied here, but this is not generalized to MPS being indispensable for every GPU calculation or to MPS being better disabled on the RTX 5080. The MPS-off state was confirmed not only from the absence of the environment variable but also from the record of the MPS server.

The strong scaling of band calculations was evaluated in a dedicated multi-node series, not mixed with the main series. With one H100 per node and 48 ranks per node, the total run time of bnc216 with cuBLAS is 252.11, 165.98, 139.22, and 187.94~s on 1, 2, 4, and 8 nodes, respectively. Four nodes give the shortest time in this range, and on eight nodes the total time increases even though the diagonalization section becomes shorter still (Appendix~\ref{app:scaling}). The node count giving the shortest time does not coincide with the condition minimizing the resource consumption, i.e.\ the time multiplied by the number of nodes.

For distributed diagonalization of the cluster, a \emph{single-binary comparison} was made for cnc384, matrix dimension 9984, with 16 ranks per node (Appendix~\ref{app:distributed}). Even on four nodes the diagonalization section with \texttt{gpusolver2} takes 225.40~s, longer than the 88.03~s of \texttt{gpusolver} in the same layout, although shorter than the 447.18~s of the CPU. This shows that keeping the single-GPU dense path is advantageous at this size and layout, but it does not imply that distributed solvers are unfavourable for larger matrices or other GPU interconnects.

\subsection{Device memory and host memory}
\label{sec:memory}

\begin{figure}[tbp]\centering
\includegraphics[width=.80\linewidth]{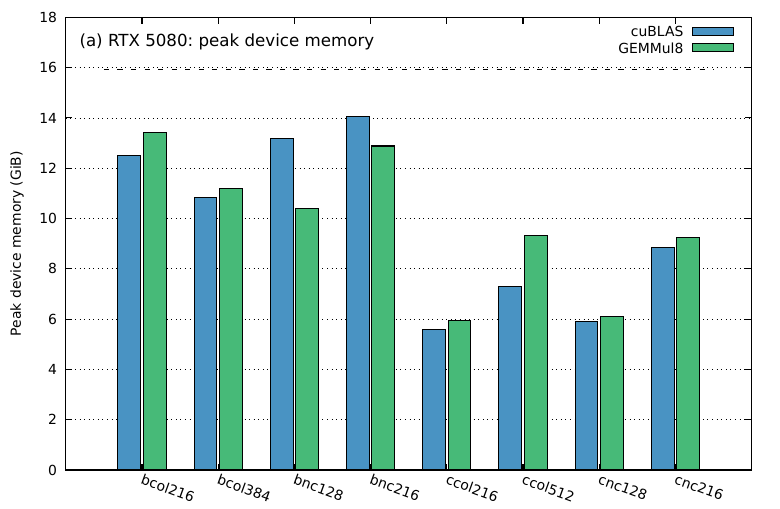}\\[2mm]
\includegraphics[width=.80\linewidth]{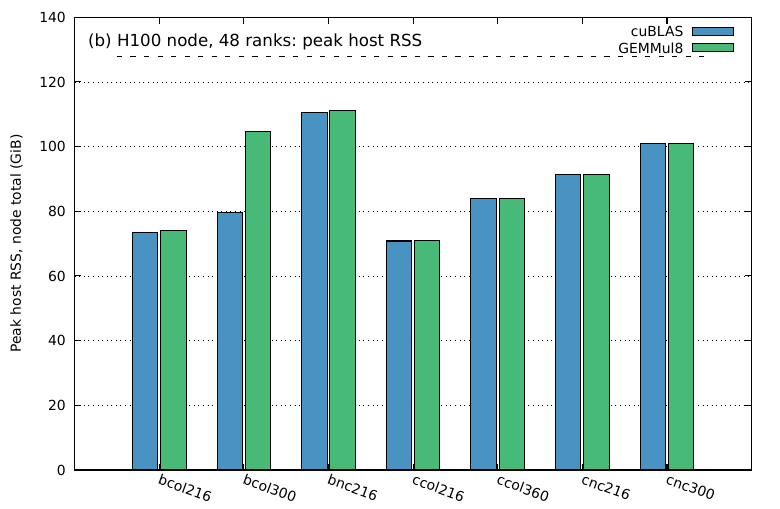}
\caption{Different constraints of GPU memory and host memory observed in the performance tests on undisplaced Si. Bars are the maxima of the high-water marks over five runs, not means or standard deviations. The dashed line in (a) is the total capacity obtained from CUDA in that environment, and in (b) the installed DDR5 capacity, which is not identical to the capacity available from the OS. Because fallbacks and the concurrency of processing change, the observed value of the GEMMul8 configuration is not always the larger.}
\label{fig:memory}
\end{figure}

Memory capacity was evaluated on the host side as well as on the GPU. Figure~\ref{fig:memory} shows, for the VRAM usage and the high-water mark of the summed RSS of node~0 recorded in the manifests, the maximum over the five runs of each configuration. The internal \texttt{mb} and \texttt{kb} fields were converted with binary units and unified to GiB in the figure. The VRAM values are those observed by the program at particular instants, and the summed RSS is also sampled; they are therefore not strict upper bounds of the instantaneous peak between samples, and the RSS may double-count shared pages.

For bnc216 on the RTX 5080, part of the GEMMul8 calls returned to cuBLAS under the prescribed free-memory reserve, but the run completed as a mixed path; this does not mean that the bnc216 calculation itself is impossible on a 16~GB-class GPU. Moreover, because of fallbacks or changes in the number of concurrent executions, the observed VRAM usage of the GEMMul8 configuration may be smaller than that of the cuBLAS configuration. The application-level peak cannot be predicted by simply adding the workspace size of a single GEMM.

For bnc216 on the H100 with 48 ranks, the maximum observed summed host RSS is 110.65~GiB with cuBLAS and 111.17~GiB with GEMMul8. An attempt at cnc384 with 48 ranks was recorded as aborted for lack of host memory and is distinguished from the successful 36-rank and 16-rank series. Owing to replicated host-side matrices and the number of MPI ranks, the host capacity can be the constraint even when the GPU VRAM has headroom. A universal ``feasible upper limit'' should therefore not be defined by the number of atoms alone; the number of basis functions, the spin representation, the number of $k$-point-owning ranks, the cell shape, and the memory-protection settings must be stated together.

\subsection{Numerical agreement and the scope of the verification}
\label{sec:accuracy}

\paragraph{Direct confirmation of practical convergence for all 24 configurations}

When all 24 configurations of displaced Si216 were re-measured with the practical convergence criterion $10^{-8}$, every run terminated by convergence before reaching the SCF cap of 200. As shown in Table~\ref{tab:practical-convergence}, the terminating iteration was 20 for ccol and 23 for bcol, bnc, and cnc. Within each path the terminating iteration coincided for CPU, cuBLAS, and GEMMul8 on both the H100 and the RTX 5080, and all values also coincided with the first-reach iterations obtained from the original stress-test histories. This result is therefore not a confirmation on a few representative series but a verification of the stopping iteration obtained by running all 24 configurations of the table in their respective environments. It is a cross-check of 48 histories, the original 24 and the present 24, not a measurement of 48 distinct physical conditions.

\begin{table}[tbp]\centering\small
\caption{Actual SCF terminating iterations in the practical-convergence re-measurement of displaced Si216 (atom~1 moved by 0.05~\AA{} along $+x$, no SOC). All 24 configurations specified \texttt{scf.criterion}=$10^{-8}$ and \texttt{scf.maxIter}=200 and were run once each. All runs terminated by convergence rather than by the cap, and each value also coincided with $i_{\rm conv}^{(8)}$ obtained from the corresponding original $10^{-13}$ history.}
\label{tab:practical-convergence}
\begin{tabular}{llccc}\toprule
GPU environment & Path & CPU & cuBLAS & GEMMul8\\\midrule
H100 & bcol & 23 & 23 & 23\\
H100 & bnc & 23 & 23 & 23\\
H100 & ccol & 20 & 20 & 20\\
H100 & cnc & 23 & 23 & 23\\\midrule
RTX 5080 & bcol & 23 & 23 & 23\\
RTX 5080 & bnc & 23 & 23 & 23\\
RTX 5080 & ccol & 20 & 20 & 20\\
RTX 5080 & cnc & 23 & 23 & 23\\
\bottomrule\end{tabular}
\end{table}

\paragraph{Energies and forces at practical convergence}

Table~\ref{tab:accuracy-practical} compares the final outputs of the practical-convergence re-measurement within the same environment and path. The differences between the GEMMul8 and cuBLAS configurations are at most $2\times10^{-12}$~Ha in the total energy and $7\times10^{-12}$~Ha/Bohr in force components. The differences between CPU and cuBLAS are at most $3.66\times10^{-10}$~Ha and $1.87\times10^{-10}$~Ha/Bohr, respectively; the CPU result is listed as a baseline of differences between distinct implementations, not as the mathematically exact solution. The table does not reinterpret the final outputs of the original stress test as values at practical convergence.

\begin{table}[tbp]\centering\small
\caption{Between-backend differences for displaced Si216 converged with the practical criterion $10^{-8}$. C: CPU, O: cuBLAS, G: GEMMul8. Energies in Ha, forces in Ha/Bohr. Energy differences are decimal differences of the printed \texttt{Utot}; forces are the maximum difference over corresponding atoms and components. Zero indicates agreement within the printed precision. $\dagger$: the G configuration includes 177 ZGEMM fallbacks, so G--O is a comparison of a mixed path against cuBLAS.}
\label{tab:accuracy-practical}
\setlength{\tabcolsep}{3pt}
\begin{tabular}{llcccc}\toprule
GPU & Path & $\Delta E_{C,O}$ & $\Delta F_{\max,C,O}$ &
$\Delta E_{G,O}$ & $\Delta F_{\max,G,O}$\\\midrule
H100 & bcol & $4.8\times10^{-11}$ & $5.6\times10^{-11}$ & $0$ & $1\times10^{-12}$\\
H100 & bnc & $1.1\times10^{-10}$ & $5.9\times10^{-11}$ & $1\times10^{-12}$ & $4\times10^{-12}$\\
H100 & ccol & $4.7\times10^{-11}$ & $1.67\times10^{-10}$ & $1\times10^{-12}$ & $1\times10^{-12}$\\
H100 & cnc & $3.66\times10^{-10}$ & $1.87\times10^{-10}$ & $1\times10^{-12}$ & $5\times10^{-12}$\\
RTX 5080 & bcol & $5.6\times10^{-11}$ & $5\times10^{-11}$ & $1\times10^{-12}$ & $1\times10^{-12}$\\
RTX 5080 & bnc$^{\dagger}$ & $1.2\times10^{-11}$ & $4.9\times10^{-11}$ & $1\times10^{-12}$ & $7\times10^{-12}$\\
RTX 5080 & ccol & $5.3\times10^{-11}$ & $1.41\times10^{-10}$ & $0$ & $2\times10^{-12}$\\
RTX 5080 & cnc & $5.3\times10^{-11}$ & $1.67\times10^{-10}$ & $2\times10^{-12}$ & $4\times10^{-12}$\\
\bottomrule\end{tabular}
\end{table}

\paragraph{Convergence and reproducibility on the mixed path}

In \path{s5a8_bnc_g11} on the RTX 5080, of the $23\times8\times7=1288$ ZGEMM requests for the 23 SCF iterations, 1111 were executed by GEMMul8 and 177 fell back to cuBLAS. The manifest recorded GEMMul8 enabled, 15 moduli, fast mode disabled, a 50\% workspace limit, a 1536~MiB free-memory reserve, MPS enabled, and termination by SCF convergence. The reason for all 177 fallbacks was the reserve policy protecting the free memory. In \path{siacc8_bnc_g11} on the H100, on the other hand, all 1288 requests were processed by GEMMul8 with no fallback. The terminating iteration was 23 in both environments, so practical convergence was confirmed directly for the mixed path on the RTX 5080 as well.

At the 22nd iteration of that bnc re-measurement on the RTX 5080, \texttt{dUele} was $7.8309\times10^{-8}$~Ha and the energy condition was not satisfied. At the 23rd iteration it became $1.019\times10^{-9}$~Ha, and the residual corresponding to the check was
\begin{equation}
 R_{23}^{\rm chk}\simeq r_{22}^{\,2}
 =(1.440320\times10^{-6})^2
 \simeq2.07452\times10^{-12}<10^{-7} .
\end{equation}
The \texttt{Uele} and printed \texttt{NormRD} of iterations 1--23 coincided with the corresponding printed values of the original 48-iteration bnc accuracy test. This coincidence is a comparison for this structure and backend configuration; it does not imply bitwise agreement of all SCF histories of the 24 configurations or invariance of the history under arbitrary fallback placements. Coincidence of the stopping iteration between a configuration using GEMMul8 and a cuBLAS-only configuration is also distinguished from exact agreement of the numbers at every iteration.

\begin{figure}[tbp]\centering
\includegraphics[width=\linewidth]{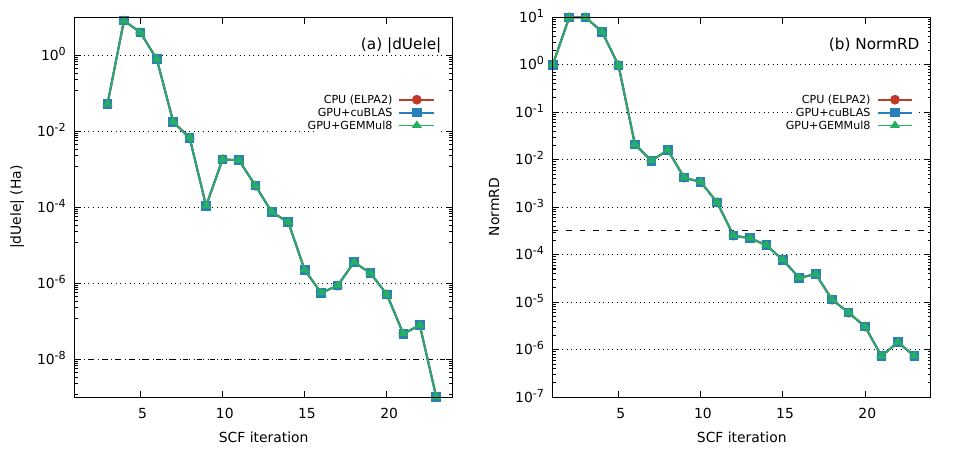}
\caption{SCF histories of bnc216 (NCL band, displaced Si216) on the RTX 5080 with 16 ranks in the practical-convergence re-measurement with threshold $10^{-8}$: (a) $|\mathtt{dUele}|$ and (b) \texttt{NormRD} at each SCF iteration (logarithmic scale). The three series, CPU (ELPA2), GPU+cuBLAS, and GPU+GEMMul8 (a mixed path including 177 ZGEMM fallbacks), nearly overlap on the plotted scale and all stop at iteration 23. Dashed lines mark the thresholds of the check ((a) $10^{-8}$~Ha; (b) $\sqrt{10^{-7}}\simeq3.16\times10^{-4}$ in printed values); the residual at iteration $i-1$ enters the stopping test at iteration $i$. Panel (a) omits the first iteration and zero differences. The history of the H100 NCL cluster with 216 atoms is shown in supplementary Fig.~\ref{fig:scfhist-h100}.}
\label{fig:scfhist}
\end{figure}

Figure~\ref{fig:scfhist} overlays the \texttt{dUele} and \texttt{NormRD} histories of the three configurations for the NCL band with 216 atoms on the RTX 5080. Even on the mixed path with fallbacks, the history overlaps those of the other two configurations iteration by iteration up to the stopping iteration. The same overlap for the fallback-free H100 NCL cluster is shown in supplementary Fig.~\ref{fig:scfhist-h100}.

\paragraph{Contrast with the stress test at threshold $10^{-13}$}
In the stress test carried out on the same 24 configurations before the re-measurement, with threshold $10^{-13}$ and a 60-iteration cap (Appendix~\ref{app:stress}), the terminating iterations scattered between 47 and 60 and included runs that stopped at the cap, yet the GEMMul8--cuBLAS differences of the final outputs remained within $10^{-12}$~Ha in the total energy and $1.3\times10^{-11}$~Ha/Bohr in force components. Moreover, when the iteration at which the conditions corresponding to $10^{-8}$ are first met is determined a posteriori from the saved histories according to the OpenMX test, it is 20 for ccol and 23 for the other three paths, coinciding with the actual terminating iterations of Table~\ref{tab:practical-convergence} for all 24 configurations. The between-threshold output differences between the re-measurement and the original stress test are also given in Appendix~\ref{app:stress}.

In the bnc216 performance test on undisplaced Si, the $25\times8\times7=1400$ ZGEMM requests for 25 SCF iterations resulted in 1207 successes and 193 fallbacks (Section~\ref{sec:rtx}); in the stress test on displaced Si216 there were 2311 successes and 377 fallbacks in 48 iterations, and in the practical-convergence re-measurement 1111 successes and 177 fallbacks in 23 iterations. These are records of separate runs with different structures or actual iteration counts and are not interchanged. The values 60 and 200 are the caps specified in the respective inputs, not the terminating iterations of those runs. For bnc on displaced Si216 on the H100, no GEMMul8 fallback occurred in either the original stress test or the re-measurement.

The matrix-level auxiliary verification used an undisplaced Si64 supercell distinct from displaced Si216, with an SCF cap of 10 and \texttt{scf.criterion}=$10^{-15}$, and the orthogonal-basis $\widetilde H$, eigenvectors $Y$, and eigenvalues saved at the third SCF iteration of a CL $\Gamma$-point calculation on the H100. With 128 occupied states and step-function occupations of two,
\begin{equation}
 K=2Y_{\rm occ}Y_{\rm occ}^{T},\qquad
 \epsilon_K=\frac{\|K_\G-K_\Oref\|_F}{\|K_\Oref\|_F}
 \label{eq:projector}
\end{equation}
were formed. This comparison does not depend on signs or orthogonal rotations within the occupied subspace. $K$ is, however, a matrix reconstructed to examine the occupied subspace in the orthogonal basis, not a direct comparison of the density matrix in the original non-orthogonal NAO basis with finite-temperature occupations.

The saved analysis gives $\epsilon_K=5.89\times10^{-15}$, $\max|\Delta K|=1.24\times10^{-14}$, and a maximum eigenvalue difference of $1.28\times10^{-15}$~Ha. Furthermore,
\begin{equation}
 r_{HK}=\frac{\|\widetilde HK-K\widetilde H\|_F}
 {\|\widetilde H\|_F\|K\|_F}
\end{equation}
rounds to $3.44\times10^{-12}$ for both configurations. This is an indicator of matrix commutativity, not the residual $\|HC-SC\varepsilon\|$ of the generalized eigenvalue problem itself. The verification is limited to this single condition at the third iteration; the same verification was not carried out for all four paths, both GPUs, or converged density matrices.

In the small auxiliary tests, MnO was treated in an antiferromagnetic CL calculation, and SOC was enabled in the NCL calculations of Cr$_2$ and GaAs; Cr$_2$ and GaAs are therefore tests under conditions different from the SOC-free Si series. In these small tests the GEMMul8 call count in the manifest was zero even with GEMMul8 enabled, because the CPU dense-matrix path for small systems was used. They are auxiliary tests of the consistency of branches and of existing results, and do not directly verify the numerical accuracy of magnetic or SOC calculations with GEMMul8. For GaAs, whose magnetization is nearly zero, differences of the printed direction angles alone are not used as an accuracy indicator either. The aggregation in Table~\ref{tab:practical-convergence} covers the 24 configurations of displaced Si216 and excludes these small auxiliary tests.

In the repeated runs under identical conditions of the main performance measurements on undisplaced Si, the printed final total energies agreed in all 50 groups. This reproducibility is agreement at the precision of the saved output; it does not guarantee bitwise agreement of all internal arrays, of the entire SCF path, or across compilers and GPUs. In summary, the results support numerical agreement within the verified conditions, but they provide neither the physical error of the DFT model itself nor an error bound of the emulation for arbitrary inputs.

\section{Discussion}

Interpreting the results requires considering the arithmetic composition, the memory, and the concurrency at the same time. First, the effect of GEMMul8 is large when the stand-alone gain over native FP64 GEMM is large and the replaced matrix products occupy a large part of the run time. The bcol384 system on the RTX 5080 satisfies this condition, whereas on the H100 the reference cuBLAS is fast and the relative share of the replaced products is correspondingly small. The difference between GPUs cannot be reduced to a single theoretical FLOPS ratio or to the matrix dimension alone.

Second, the GEMMul8 workspace involves a trade-off between performance and capacity. Even at the prescribed accuracy setting, memory protection can turn the path into one mixed with cuBLAS; on the other hand, merely relaxing the workspace limit to reduce fallbacks may squeeze resident arrays or the concurrency of other ranks. As bcol300 on the H100 shows, a small gain in the diagonalization can coexist with a deterioration of the total. In practice, therefore, not only the number of successful GEMMul8 calls but the total run time and the memory placement must be evaluated.

Third, what our implementation changes is the explicit GEMMs outside cuSOLVER. Its scope of modification and its performance ceiling differ from those of mixed-precision methods that change the algorithm including the interior of the eigensolver~\cite{woo2026,higham2022}. The advantage of our scheme is that the existing FP64 data representation and solution path are retained and the numerical differences caused by the switch can be tracked within a limited scope; this does not, however, amount to an automatic accuracy guarantee, and verification according to the condition number and the input distribution remains necessary~\cite{uchino-error}.

The re-measurement of all 24 configurations of displaced Si216 with the practical convergence criterion terminated by convergence in every case, with identical terminating iterations for CPU, cuBLAS, and GEMMul8 within each path. Runs that did not reach the original $10^{-13}$ termination test should therefore not be treated wholesale as unconverged in the practical sense. Separating the between-backend difference at practical convergence from the between-threshold difference for the same backend distinguishes the numerical difference due to switching the implementation from the numerical change due to the different stopping point. The GEMMul8--cuBLAS difference at practical convergence is at most $2\times10^{-12}$~Ha in the total energy and $7\times10^{-12}$~Ha/Bohr in force components, whereas for the same configuration the energy and forces at the point of convergence with threshold $10^{-8}$ (iteration 20 or 23) differ from those at the end of the $10^{-13}$ stress test (iterations 47--60) by at most about $1.22\times10^{-8}$~Ha and $1.83\times10^{-8}$~Ha/Bohr. The latter is the difference between stopping at the $10^{-8}$ test and continuing the iteration, and its maximum exceeds the corresponding maximum between-backend difference by more than three orders of magnitude for both energy and forces. These are results for the present test set and do not derive an error guarantee for arbitrary physical quantities from the mere satisfaction of a stopping condition.

The limitations of this work include the Si-centred test set, the limited bases and $k$ points, the difference between fixed-work measurements and time to convergence, and the constraint of the printed precision. Since SOC was not included in the Si NCL calculations and the SOC-enabled auxiliary tests on Cr$_2$ and GaAs involved no GEMMul8 calls, the performance and numerical agreement of GEMMul8 paths in calculations with SOC lie outside the directly verified scope. Verification for general magnetic metals, for strongly different exponent distributions, and for ill-conditioned overlap matrices also remains. The practical-convergence re-measurement covers all 24 configurations but with one run each, and is not a repeated timing measurement of the statistical speedup of the time to convergence; the confirmation of accuracy and terminating iterations at practical convergence is therefore not confused with the speedups of the fixed-25-iteration performance measurements. Fixed-work comparisons that include early-terminating cases such as the 64-atom systems would require a measurement design that controls the iteration count independently of the termination condition.

Future work includes the selection of the number of moduli and error evaluation according to the input values~\cite{uchino-error}, extension to larger sizes on GPUs with more VRAM, the FP8 backend~\cite{uchino-fp8}, and deployment to other GPU environments; no performance was extrapolated to unmeasured GPUs in this work. Since neither integrated power consumption nor system cost was compared, no quantitative advantage in energy efficiency or cost effectiveness is concluded from these measurements alone.

\section{Conclusions}

We have integrated GEMMul8 into the four dense-matrix computational paths of OpenMX 4.0 GPU and evaluated the additional effect of matrix-multiplication emulation by comparison with the cuBLAS FP64 configuration on the same GPU. The NCL series of undisplaced Si used for the performance evaluation performed spinor calculations without SOC. For bcol384 on the RTX 5080, at a fixed 25 SCF iterations, the diagonalization section is accelerated by 2.11 and the total run time by 1.81; ccol512 also gains 1.53 in the diagonalization section. On the H100 main series, on the other hand, the diagonalization speedup stays within 0.978--1.019, and for bcol300 the total time increases by about 15\%, consistent with an interaction between workspace allocation and other processing.

A model based on stand-alone GEMMs and call counts reproduces the diagonalization speedups within 1.5\% for five of the eight RTX 5080 conditions, with residuals that include several run-time factors. The re-measurement of all 24 configurations of displaced Si216 with \texttt{scf.criterion}=$10^{-8}$ and a 200-iteration cap terminated by convergence in every case, at iteration 20 for ccol and 23 for bcol, bnc, and cnc; within each path the iterations coincided for CPU, cuBLAS, and GEMMul8 in both environments, and the first-reach iterations from the original $10^{-13}$ histories were corroborated for all configurations. The differences between the GEMMul8 and cuBLAS configurations at practical convergence are at most $2\times10^{-12}$~Ha in the total energy and $7\times10^{-12}$~Ha/Bohr in force components. For bnc on the RTX 5080 we confirmed practical convergence of a mixed path with 177 fallbacks, and on the H100 the convergence of the same bnc path without fallbacks. These results show numerical agreement in the practically converged state, separately from the final-output comparison of the tight $10^{-13}$ stress test. The scopes of the performance tests on undisplaced Si, the accuracy tests on displaced Si216, and the SOC-enabled small auxiliary tests are kept distinct. In conclusion, GEMMul8 is an effective option for accelerating NAO-based DFT on GPUs whose native FP64 throughput is relatively low, and deciding on its use requires confirmation not only of stand-alone arithmetic speed but of the total run time, the convergence conditions, and the actual execution path.

\FloatBarrier
\appendix

\section{Stress test at threshold $10^{-13}$: conditions, history analysis, and results}\label{app:stress}

The stress test that preceded the re-measurement used \texttt{scf.criterion}=$10^{-13}$~Ha and a 60-iteration cap for the same 24 configurations. In the authors' experience with OpenMX, stably satisfying the termination condition with \texttt{scf.criterion} below $10^{-12}$~Ha is very difficult. The value $10^{-13}$~Ha is therefore not a recommendation for regular calculations; it was chosen deliberately as a stress condition under which the iteration continues beyond practical convergence, so that the SCF histories and the final numerical differences of energies and forces can be examined in a regime where rounding errors may become visible. It was not assumed that all runs would satisfy this threshold, nor is $10^{-12}$~Ha regarded as a mathematical convergence limit common to arbitrary systems.

Both the stress test and the practical-convergence re-measurement of displaced Si216 used \texttt{scf.Mixing.Type=rmm-diisk} in fixed-basis calculations without orbital optimization. In \path{source/DFT.c} of the measurement tag \texttt{v2.0\_thesis} (commit \texttt{db72cd4}), the \texttt{Cnt\_switch=0} branch tests both the energy difference and the density residual~\cite{openmxgpu-source}. Writing $\epsilon$ for the threshold, $U_i$ for \texttt{Uele} at iteration $i$, and $R_i^{\rm chk}$ for the internal variable \texttt{NormRD[0]} at the check, the convergence condition is
\begin{equation}
 i\geq2,\qquad
 |U_i-U_{i-1}|<\epsilon,\qquad
 R_i^{\rm chk}<10\epsilon .
 \label{eq:scf-stop}
\end{equation}
The comparisons are expressed in the same atomic-unit numbers as in the code. \texttt{dUele} is the absolute difference of \texttt{Uele}, not of \texttt{Utot}, between consecutive iterations, and 1.0 is assigned at the first iteration.

In this mixing path, the density mixing is performed after the convergence check; the value printed afterwards as \texttt{NormRD} is the square root of \texttt{fabs(NormRD[0])}. Writing $r_i$ for the printed residual of iteration $i$, the internal value at the check corresponds, within the printed rounding, to $R_i^{\rm chk}\simeq r_{i-1}^{\,2}$. The printed \texttt{NormRD} of the same iteration must therefore not be compared directly with \texttt{scf.criterion}. The first reach of the practical convergence criterion determined from the saved histories of the original stress test is defined as
\begin{equation}
 i_{\rm conv}^{(8)}
 =\min\left\{i\geq2:
 |U_i-U_{i-1}|<10^{-8},\quad r_{i-1}^{\,2}<10^{-7}\right\},
 \label{eq:practical-convergence}
\end{equation}
where the residual condition corresponds to $r_{i-1}<\sqrt{10^{-7}}\simeq3.1623\times10^{-4}$ in the previous printed value. Even if only the energy difference becomes small in an early iteration, it is not counted as convergence unless the density-residual condition is also met.

\paragraph{Between-threshold differences}

For the same environment, path, and backend, the final output of the practical-convergence re-measurement was compared with the final output of the original $10^{-13}$ stress test. Over the 24 pairs, the between-threshold difference is at most about $1.22\times10^{-8}$~Ha in the total energy and about $1.83\times10^{-8}$~Ha/Bohr in force components. This comparison is distinct from the between-backend differences of Table~\ref{tab:accuracy-practical} and quantifies the numerical change due to the different stopping point. Because some original runs did not reach the tight threshold, this difference is not regarded as an error bound with respect to a fully converged solution, nor is it generalized to an accuracy guarantee at the $10^{-8}$ criterion for arbitrary physical quantities.

In a separately checked \texttt{sidia333\_cluster} input of undisplaced Si216, a calculation specifying $10^{-8}$ terminated after 15 iterations. This input does not displace atom~1 and the coordinates are printed with a different number of digits, so it is distinguished from a rerun of the identical input that would verify the 20 iterations of displaced Si216; identical iteration counts cannot be demanded merely because the number of atoms and the solver are the same.

\paragraph{Contrast with the tight-threshold stress test}

Table~\ref{tab:accuracy} compares the final physical quantities of the original $10^{-13}$ stress test on displaced Si216. The differences between the GEMMul8 configuration (including the mixed path caused by fallbacks) and the cuBLAS configuration are at most $10^{-12}$~Ha in the total energy and $1.3\times10^{-11}$~Ha/Bohr in force components; the CPU-versus-cuBLAS total-energy difference in the same table is at most $1.79\times10^{-10}$~Ha. These are the final outputs of the original runs against the tight input threshold and are distinguished from Table~\ref{tab:accuracy-practical}, which reports the practical-convergence re-measurement.

\begin{table}[tbp]\centering\small
\caption{Differences of the final outputs for displaced Si216 (atom~1 moved by 0.05~\AA{} along $+x$; no SOC, including the NCL paths). C: CPU, O: cuBLAS, G: GEMMul8. The SCF counts are the original terminating iterations at the input threshold $10^{-13}$; $*$ marks runs that did not satisfy that tight test and reached the 60-iteration cap. The terminating iterations and quantities of the same 24 configurations re-measured with the practical criterion are given in Tables~\ref{tab:practical-convergence} and \ref{tab:accuracy-practical}. $\dagger$: the G configuration terminated after 48 iterations and includes 377 ZGEMM fallbacks, so the G--O differences of this row compare a mixed path with cuBLAS. Energies in Ha, forces in Ha/Bohr. Values are decimal differences of printed values; zero means agreement within the printed precision.}
\label{tab:accuracy}
\setlength{\tabcolsep}{3pt}
\begin{tabular}{llcccc}\toprule
GPU & Path & SCF C/O/G & $\Delta E_{C,O}$ & $\Delta E_{G,O}$ & $\Delta F_{\max,G,O}$\\\midrule
H100 & bcol & $54/58/56$ & $4.9\times10^{-11}$ & $0$ & $3\times10^{-12}$\\
H100 & bnc & $60^{*}/47/47$ & $1\times10^{-11}$ & $1\times10^{-12}$ & $0$\\
H100 & ccol & $60^{*}/60^{*}/55$ & $4.6\times10^{-11}$ & $1\times10^{-12}$ & $8\times10^{-12}$\\
H100 & cnc & $60^{*}/51/56$ & $1.18\times10^{-10}$ & $1\times10^{-12}$ & $1.3\times10^{-11}$\\
RTX 5080 & bcol & $60/60^{*}/60^{*}$ & $5\times10^{-11}$ & $0$ & $0$\\
RTX 5080 & bnc$^{\dagger}$ & $60^{*}/52/48$ & $3.4\times10^{-11}$ & $1\times10^{-12}$ & $3\times10^{-12}$\\
RTX 5080 & ccol & $58/55/55$ & $4.5\times10^{-11}$ & $0$ & $1\times10^{-12}$\\
RTX 5080 & cnc & $60^{*}/60^{*}/60^{*}$ & $1.79\times10^{-10}$ & $1\times10^{-12}$ & $0$\\
\bottomrule\end{tabular}
\end{table}

Table~\ref{tab:accuracy} contains runs that did not reach the tight input criterion $10^{-13}$, but this does not mean that they failed to reach the practical criterion $10^{-8}$. For ccol216 on the H100, for example, CPU and cuBLAS stopped at the 60-iteration cap while GEMMul8 satisfied the input test at iteration 55, yet all three configurations reached the practical criterion at iteration 20; for cnc216 on the RTX 5080, all three configurations failed the input criterion at 60 but reached the practical criterion at 23. Hence satisfying the practical criterion in all 24 configurations is distinguished from satisfying the $10^{-13}$ termination condition of the input in all of them. Small numerical differences accompanying the switch to GEMMul8 can appear as differences in the iteration count under an extremely tight termination test. In the bnc accuracy test of displaced Si216 on the RTX 5080, the GEMMul8 configuration terminated after 48 SCF iterations with $48\times8\times7=2688$ ZGEMM requests, of which 2311 succeeded in GEMMul8 and 377 fell back to cuBLAS; the $\Delta E_{G,O}$ and $\Delta F_{\max,G,O}$ of that row of Table~\ref{tab:accuracy} therefore represent \emph{a mixed GEMMul8/cuBLAS path versus cuBLAS}, not a path in which all target products were processed by GEMMul8.

\section{Main H100 measurements}
\label{app:h100}

Table~\ref{tab:h100} lists the total run times of the seven main H100 conditions on undisplaced Si for 48 ranks on one node and 25 SCF iterations, together with the additional speedups of GEMMul8. The columns CPU, cuBLAS, and GEMMul8 give the mean and sample standard deviation of the total run time (\texttt{Max\_Time} of the whole run) over three CPU runs and five GPU runs each. The same executable was used for GPU and CPU, the comparison being made by switching to the CPU path, in which the dense eigenvalue problems are solved by ELPA2 on the CPU cores. $S_\G^{\rm diag}$ and $S_\G^{\rm total}$ are the ratios of the mean cuBLAS time to the mean GEMMul8 time for the diagonalization section and for the total run; the diagonalization times themselves are tabulated in \path{production_summary.csv} (Appendix~\ref{app:provenance}), and those of bcol300 are shown in Fig.~\ref{fig:phases}.

Three points can be read from the table. First, the CPU-to-GPU speedup of the total run time, the ratio of the CPU column to the cuBLAS column, ranges from 1.70 for ccol216 to 4.16 for cnc300: it is 2.95--3.75 for the band paths and 3.20--4.16 for the NCL cluster, but only 1.70--2.01 for the CL cluster, in which the dense part is a small share of the total and only that part runs on the single GPU while the remaining operations are distributed over the 48 ranks. Second, replacing cuBLAS by GEMMul8 changes the diagonalization section by about 2\% at most in either direction ($S_\G^{\rm diag}=0.978$--$1.019$), in agreement with the stand-alone measurements of Section~\ref{sec:micro}, where the H100 cuBLAS/GEMMul8 time ratios of the application shapes are at most 1.55 and, aggregated over the calls of a run, below one for several conditions. Third, the total-run-time ratios lie within 0.986--1.006 except for bcol300, where the total time increases by 15\% ($S_\G^{\rm total}=0.870$) although the diagonalization section becomes slightly shorter; this case, in which \texttt{Set\_Hamiltonian} and \texttt{Set\_Density\_Grid} become longer and the device memory use rises, is analysed in Section~\ref{sec:h100}. The larger standard deviation of the GEMMul8 run time of bcol300 ($\pm6.39$~s against $\pm1.76$~s for cuBLAS) accompanies the same interaction.

\begin{table}[htbp]\centering\small
\caption{Total run times [s] of undisplaced Si in the H100 environment (mean $\pm$ sample standard deviation) and additional speedups of GEMMul8. $S_\G^{\rm diag}$ refers to the diagonalization section and $S_\G^{\rm total}$ to the total run time.}
\label{tab:h100}\setlength{\tabcolsep}{3.3pt}
\begin{tabular}{lccccc}\toprule
System & CPU & cuBLAS & GEMMul8 & $S_\G^{\rm diag}$ & $S_\G^{\rm total}$\\\midrule
bcol216 & $\meanpm{240.48}{0.75}$ & $\meanpm{81.45}{0.87}$ & $\meanpm{81.33}{1.77}$ & 1.008 & 1.002\\
bcol300 & $\meanpm{489.47}{1.53}$ & $\meanpm{131.37}{1.76}$ & $\meanpm{151.03}{6.39}$ & 1.017 & 0.870\\
bnc216 & $\meanpm{945.85}{2.70}$ & $\meanpm{252.14}{1.46}$ & $\meanpm{255.82}{1.64}$ & 0.978 & 0.986\\
ccol216 & $\meanpm{87.45}{0.55}$ & $\meanpm{51.35}{1.28}$ & $\meanpm{51.90}{0.73}$ & 1.019 & 0.989\\
ccol360 & $\meanpm{139.97}{0.54}$ & $\meanpm{69.56}{0.83}$ & $\meanpm{69.53}{0.31}$ & 0.987 & 1.000\\
cnc216 & $\meanpm{321.99}{3.77}$ & $\meanpm{100.52}{3.11}$ & $\meanpm{99.91}{0.77}$ & 1.008 & 1.006\\
cnc300 & $\meanpm{563.80}{6.60}$ & $\meanpm{135.52}{0.71}$ & $\meanpm{135.13}{1.52}$ & 1.003 & 1.003\\
\bottomrule\end{tabular}
\end{table}

\section{Strong scaling of band calculations}
\label{app:scaling}

Table~\ref{tab:scaling} shows the node-count dependence of two band calculations on undisplaced Si, bcol300 and bnc216, measured in a dedicated strong-scaling series separate from the main series of Table~\ref{tab:h100}; the one-node values are also taken from this series so that measurement sets are not mixed (they agree with Table~\ref{tab:h100} within the standard deviations, 131.27 against 131.37~s and 252.11 against 252.14~s). Each node contributes 48 MPI ranks and one H100, so the node count multiplies CPU cores, host memory, and GPUs together; it is not an experiment that adds GPUs only. The cuBLAS configuration and 25 SCF iterations were used with five or seven runs per point, and the table gives the mean and sample standard deviation of the total run time and of the diagonalization section, the speedup $S_p=\overline T_1/\overline T_p$ of the total time relative to one node, and the parallel efficiency $E_p=S_p/p$.

The diagonalization section decreases monotonically with the node count, from 83.05 to 22.32~s for bcol300 and from 154.29 to 55.06~s for bnc216, because the eight computed $k$ points are distributed over the $k$-point-owning ranks of the participating nodes and each GPU processes fewer of them. The remainder of the run time, the difference between the total and the diagonalization, behaves differently: it is 48.2, 40.0, 37.7, and 51.8~s for bcol300 and 97.8, 72.9, 73.6, and 132.9~s for bnc216 on 1, 2, 4, and 8 nodes, so that it shrinks only up to four nodes and grows again on eight. This part was not profiled further in this series, and the increase on eight nodes is reported without attributing it to a particular operation. As a result the total time is shortest on four nodes ($S_4=1.93$ and $1.81$, $E_4=48\%$ and $45\%$) and longer again on eight ($S_8=1.77$ and $1.34$). The node count that minimizes the time therefore differs from the one that minimizes the resource consumption, the time multiplied by the number of nodes, which is smallest on one node. These results refer to the fixed system sizes and the eight $k$ points used here and are not extrapolated to other $k$-point sets or to the cluster paths.

\begin{table}[htbp]\centering\small
\caption{Node-count dependence of H100 band calculations on undisplaced Si. 48 ranks and one GPU per node, cuBLAS, mean $\pm$ sample standard deviation. $S_p=\overline T_1/\overline T_p$ and $E_p=S_p/p$.}
\label{tab:scaling}
\begin{tabular}{lcccccc}\toprule
System & Nodes & Runs & Total [s] & Diagonalization [s] & $S_p$ & $E_p$ [\%]\\\midrule
bcol300 & 1 & 5 & $\meanpm{131.27}{1.64}$ & $\meanpm{83.05}{1.25}$ & 1.00 & 100.0\\
bcol300 & 2 & 5 & $\meanpm{87.44}{0.92}$ & $\meanpm{47.45}{0.83}$ & 1.50 & 75.1\\
bcol300 & 4 & 7 & $\meanpm{68.16}{1.29}$ & $\meanpm{30.42}{1.14}$ & 1.93 & 48.1\\
bcol300 & 8 & 7 & $\meanpm{74.11}{1.17}$ & $\meanpm{22.32}{0.85}$ & 1.77 & 22.1\\
bnc216 & 1 & 5 & $\meanpm{252.11}{0.65}$ & $\meanpm{154.29}{0.95}$ & 1.00 & 100.0\\
bnc216 & 2 & 5 & $\meanpm{165.98}{2.36}$ & $\meanpm{93.12}{1.22}$ & 1.52 & 75.9\\
bnc216 & 4 & 5 & $\meanpm{139.22}{3.54}$ & $\meanpm{65.58}{2.75}$ & 1.81 & 45.3\\
bnc216 & 8 & 7 & $\meanpm{187.94}{3.54}$ & $\meanpm{55.06}{1.12}$ & 1.34 & 16.8\\
\bottomrule\end{tabular}
\end{table}

\section{Single-binary comparison of cluster solvers}
\label{app:distributed}

Table~\ref{tab:distributed} compares, for the NCL cluster cnc384 on undisplaced Si (matrix dimension $n_2=9984$), three ways of solving the dense generalized eigenvalue problem on one to four H100 nodes with 16 ranks per node, 25 SCF iterations, and three runs each: the CPU path, in which ELPA2 solves the problem on the CPU cores of all nodes; \texttt{gpusolver}, the path used in the main measurements, in which the dense-matrix part is processed on a single GPU while the remaining operations are distributed with MPI; and \texttt{gpusolver2}, the reference path in which the dense matrices themselves are distributed over the GPUs of the nodes with ELPA GPU and COSMA. All configurations used the same binary of the pre-freeze commit \texttt{f5fb5b3}, and GEMMul8 was enabled in the \texttt{gpusolver} measurements, so the table compares implementation strategies and is not a GEMMul8 on/off comparison alone. The table lists the diagonalization section; the total run times of the same runs are 1750.8, 184.7, and 667.1~s on one node and 560.8, 127.8, and 262.0~s on four nodes for the CPU path, \texttt{gpusolver}, and \texttt{gpusolver2}, respectively, and follow the same ordering.

The three paths scale differently. The CPU path speeds up almost in proportion to the node count (1342.48 to 447.18~s, a factor of 3.0 on four nodes), as expected for a distributed CPU eigensolver. \texttt{gpusolver} changes little (94.70 to 88.03~s), because the dense part is executed on one GPU whatever the node count and only the distributed remainder of the section benefits from additional nodes. \texttt{gpusolver2} scales (566.45 to 225.40~s, a factor of 2.5) but remains 2.6 times slower than \texttt{gpusolver} on four nodes, and the measured range does not reach a crossover between the two. At this matrix size and layout, keeping the single-GPU dense path is therefore advantageous; the measurements do not show how the comparison develops for larger matrices, more nodes, or GPU interconnects other than the PCIe nodes used here.

\begin{table}[htbp]\centering\small
\caption{Diagonalization section [s] of cnc384 on undisplaced Si (mean $\pm$ sample standard deviation). Unified to 16 ranks per node; measurements with different binaries or rank densities were not mixed.}
\label{tab:distributed}
\begin{tabular}{cccc}\toprule
Nodes & CPU ELPA2 & \texttt{gpusolver} & \texttt{gpusolver2}\\\midrule
1 & $\meanpm{1342.48}{2.04}$ & $\meanpm{94.70}{0.99}$ & $\meanpm{566.45}{3.95}$\\
2 & $\meanpm{749.96}{2.41}$ & $\meanpm{90.08}{0.10}$ & $\meanpm{358.79}{0.65}$\\
3 & $\meanpm{552.81}{0.46}$ & $\meanpm{89.05}{0.91}$ & $\meanpm{272.08}{0.83}$\\
4 & $\meanpm{447.18}{1.73}$ & $\meanpm{88.03}{0.21}$ & $\meanpm{225.40}{1.63}$\\
\bottomrule\end{tabular}
\end{table}

\section{Provenance of the measurement series and correspondence to figures and tables}
\label{app:provenance}

The main performance measurements on undisplaced Si, the accuracy tests on displaced Si216, and the band strong-scaling measurements on undisplaced Si used executables built from \texttt{v2.0\_thesis} (\texttt{db72cd4}). The MD5 sums of the main binaries of the two environments are
\begin{quote}\small
H100: \texttt{11227640dc6f8a8b194ddcd9ab811917}\\
RTX 5080: \texttt{e3e64c35f48a9058cd98b5e923810ae3}
\end{quote}
The expression ``single-binary comparison'' is used within each environment and series; identical executables across builds for different GPUs are not required.

The supplementary 36-rank cnc384 cuBLAS/GEMMul8 comparison used the \texttt{e227507} binary and the 16-rank series of Table~\ref{tab:distributed} the \texttt{f5fb5b3} binary; the measurement archive records these as direct ancestors of the measurement commit \texttt{db72cd4b}, 16 and 8 commits behind it, respectively. They are treated separately from the main series and are not presented as if they formed a continuous all-size series with the main binary. The aborted 48-rank series and the runs excluded by the slow-node convention are also preserved separately from the retained set.

The main tables and Figs.~\ref{fig:layer1}--\ref{fig:phases}, \ref{fig:mps}, and \ref{fig:memory} correspond to \path{work/reaggregate/production_runs.csv} and \path{work/reaggregate/production_summary.csv}, re-aggregated from the individual \texttt{.out} files and manifests by \path{work/reaggregate/reaggregate.py}. Table~\ref{tab:amdahl} and Fig.~\ref{fig:amdahl} correspond to \path{work/gemm_micro/amdahl_5080.csv} and \path{work/gemm_micro/amdahl_h100.csv}, written by the analysis scripts kept with the stand-alone tests; their inputs are the stand-alone times of \path{gemm_micro_5080.csv} and \path{gemm_micro_h100.csv}, the call counts listed in \path{work/reaggregate/amdahl_sites.csv}, and the unrounded mean times of the retained runs in \path{production_runs.csv}. The four cnc128 stand-alone conditions were measured on 1 November 2026, after revision 4 of the archive had been frozen, with the same \texttt{bench\_gemm} program against a GEMMul8 library of the same commit built from a fresh checkout of commit \texttt{db72cd4b}, then the target of the \texttt{v2.0\_thesis} tag; their times (\path{work/gemm_micro/gemm_micro_5080_cnc128.csv}, also merged into \path{gemm_micro_5080.csv}), the call counts from the cnc128 manifests, and the build record (\path{work/gemm_micro/gemm_micro_5080_cnc128_result.txt} and \path{bench_env_5080_cnc128.txt}) are included in the archive from revision 5 on. Table~\ref{tab:accuracy} corresponds to \path{work/reaggregate/accuracy_comparison.csv}, which holds the decimal differences, and \path{work/reaggregate/accuracy_runs.csv}, which holds the SCF counts of the individual outputs and, as \texttt{converged\_by\_criterion}, the \texttt{scf\_converged} flag of their manifests on which the asterisks of the table are based. The supplementary strong-scaling and distributed-solver results are separated into \path{work/reaggregate/strong_scaling_runs.csv} and \path{work/reaggregate/distributed_solver_runs.csv}, which name the retained raw logs; every other run directory of the archive is listed in \path{work/reaggregate/supplementary_series_runs.csv}. The re-aggregation script records the SHA-256 of the files it processes in \path{work/reaggregate/reaggregate_sha256.txt}.

Tables~\ref{tab:practical-convergence} and \ref{tab:accuracy-practical} correspond to the 24 practical-convergence re-measurement runs executed on 8 September 2026. On the H100 the run names are \texttt{siacc8\_} followed by the path name (\texttt{bcol}, \texttt{bnc}, \texttt{ccol}, \texttt{cnc}) and the configuration identifier (\texttt{\_c11}, \texttt{\_o11}, \texttt{\_g11}); on the RTX 5080 they are \texttt{s5a8\_} followed by the same path names and \texttt{\_c1611}, \texttt{\_o11}, \texttt{\_g11}. These are distinguished from the prefixes \texttt{siacc\_} and \texttt{s5a\_} of the original stress test. The aggregations are \path{work/siacc8_thesis_result.txt} and \path{work/s5a8_thesis_result.txt}, and the input, standard output, result output, manifest, and environment record of each run are bundled.

The binaries of the re-measurement are the same \texttt{v2.0\_thesis} builds as in the original measurements of each environment, with MD5 sums identical to the H100 and RTX 5080 values above. The twelve H100 runs correspond to jobs 984985--984996. The SCF cap was 200 and the threshold $10^{-8}$ in all 24 runs; the records of the original $10^{-13}$, 60-iteration runs were not overwritten.

The convergence test and the order of density mixing and output are based on the frozen sources \path{source/DFT.c} and \path{source/outputfile1.c} of the measurement tag~\cite{openmxgpu-source}; the reference commit is \texttt{db72cd4b\allowbreak 6a3a6275\allowbreak 1f12a872\allowbreak 3184f401\allowbreak 47537029}, and neither an updating default branch nor a distributed copy on Google Drive is used as the identifier of the source version. The simplified energy-difference auxiliary indicator included in the aggregation is distinguished from the OpenMX stopping test defined in Eq.~\eqref{eq:scf-stop}.

The new bnc manifests are \path{s5a8_bnc_g11.manifest.json} on the RTX 5080 and \path{siacc8_bnc_g11.manifest.json} on the H100. Both record $10^{-8}$, 23 iterations, a 50\% workspace limit, and a 1536~MiB free-memory reserve, but the ZGEMM successes/fallbacks were 1111/177 on the RTX 5080 and 1288/0 on the H100. The record of 48 iterations and 2311/377 in the old \path{s5a_bnc_g11.manifest.json} corresponds to the original stress test only. The CPU fallback of \texttt{Set\_Hamiltonian} is also distinguished from the GEMM fallback from GEMMul8 to cuBLAS.

The direct verification of convergence in this paper uses the 24 runs with the new names above; two earlier preliminary additional calculations of ccol and bnc with the same input names are not added to the iteration counts of these 24 runs.

\FloatBarrier

\subsection*{Supplementary figure: SCF history of the H100 NCL cluster}
\begin{figure}[tbp]\centering
\includegraphics[width=\linewidth]{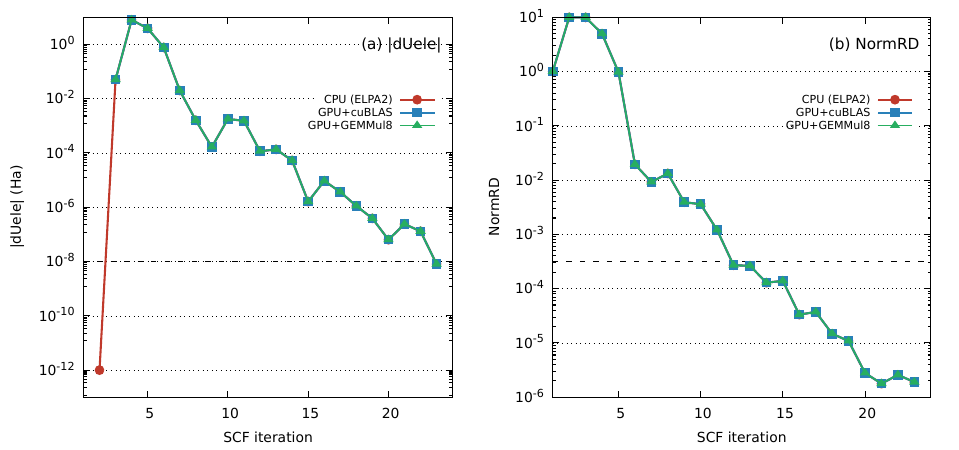}
\caption{Supplementary figure: SCF histories of cnc216 (NCL cluster, displaced Si216) on the H100 with 48 ranks in the practical-convergence re-measurement with threshold $10^{-8}$: (a) $|\mathtt{dUele}|$ and (b) \texttt{NormRD} at each SCF iteration (logarithmic scale). The three series, CPU (ELPA2), GPU+cuBLAS, and GPU+GEMMul8 (no fallbacks), nearly overlap on the plotted scale and all stop at iteration 23. Dashed lines mark the thresholds of the check ((a) $10^{-8}$~Ha; (b) $\sqrt{10^{-7}}$ in printed values); the residual at iteration $i-1$ enters the stopping test at iteration $i$. Panel (a) omits the first iteration and zero differences.}
\label{fig:scfhist-h100}
\end{figure}

Figure~\ref{fig:scfhist-h100} overlays $|\mathtt{dUele}|$ and \texttt{NormRD} at each SCF iteration for the three configurations CPU (ELPA2), GPU+cuBLAS, and GPU+GEMMul8 in the H100 NCL cluster with 216 atoms (practical-convergence re-measurement, threshold $10^{-8}$). In this path the GEMMul8 configuration had no DGEMM or ZGEMM fallbacks, so all twelve DGEMMs and the single ZGEMM back-transformation per iteration were executed by emulation. The three histories nearly overlap on the plotted scale and, like the RTX 5080 NCL band (Fig.~\ref{fig:scfhist}), stop at iteration 23. The data and the plotting script are provided as ancillary files of this article (\path{data/f7_h100_cnc.dat} and \path{fig8_scf_history_h100.plt}).

\begin{acknowledgments}

We thank the developers of GEMMul8 for enabling this study. This research in part used the computational resources of Pegasus provided by the general use program of the Center for Computational Sciences, University of Tsukuba, and also used the computational resources of the Research Center for Computational Science, Okazaki, for the development and debugging of OpenMX 4.0 GPU v2.0. This work was partially supported by JSPS KAKENHI Grant Number 25K07201. One of the authors (Kawai) received assistance from Claude and ChatGPT in language editing and proofreading of the manuscript and in checking the consistency between the text and the measurement archive; the authors are responsible for the content and any remaining errors.

\end{acknowledgments}

\section*{Program Summary}

\begin{description}[style=nextline,leftmargin=1em,labelindent=0pt,labelwidth=0pt,labelsep=0pt]
\item[Program title] OpenMX 4.0 GPU (v2.0).
\item[Developer's repository] \url{https://github.com/dc1394/openmx4.0_gpu}. The source of the main measurements is \texttt{v2.0\_thesis} (\texttt{db72cd4})~\cite{openmxgpu-source}.
\item[Licensing provisions] OpenMX GPU is GPL-3.0-or-later; GEMMul8 is MIT; other bundled components retain their own licenses.
\item[Programming languages] C, C++/CUDA, OpenACC (the CPU libraries include Fortran).
\item[External routines] cuBLAS, cuSOLVER, GEMMul8 v3.2.0, OpenMPI, FFTW. The regular CPU ELPA path uses the bundled ELPA 2018.05; \texttt{gpusolver2} uses ELPA 2026.02 and COSMA 2.8.4 with their dependencies.
\item[Nature of problem] GPU acceleration of SCF calculations, including the generalized eigenvalue problem, in NAO-basis Kohn--Sham DFT.
\item[Solution method] Standardization with the overlap matrix and dense eigenvalue solution. The explicit real and complex matrix products are switched to FP64 emulation by INT8 matrix operations based on Ozaki scheme II.
\item[Restrictions] The evaluation in this paper covers the four computational paths in an NVIDIA CUDA environment. The performance tests used undisplaced Si, the verification of energies, forces, and the practical convergence criterion used displaced Si216, and the matrix-level auxiliary verification used undisplaced Si64, none of which included SOC even in the NCL calculations. The SOC-enabled Cr$_2$ and GaAs cases are auxiliary tests of the small-system CPU dense-matrix path, not a direct verification of GEMMul8 paths with SOC. Fallback to cuBLAS is permitted under workspace constraints. No accuracy guarantee for arbitrary inputs with 15 moduli and no performance guarantee on unmeasured GPUs are included.
\end{description}

\section*{Data availability}

The code is maintained in the repository above. The executables of all runs were built from commit \texttt{db72cd4b}, the freeze commit on which the tag \texttt{v2.0\_thesis} was created and which this paper cites as \texttt{v2.0\_thesis} (\texttt{db72cd4}). On 2 November 2026 the tag was moved to commit \texttt{7af14b87} of the branch \texttt{release/v2.0\_thesis}, which adds the measurement results, the archive, and the supplement described below; its \path{source/}, \path{doc/}, and \path{DFT_DATA19/} trees are identical to those of \texttt{db72cd4b}, which remains in the history of the branch. The measurement archive referred to in this paper is revision 7 of \path{openmx40gpu_v2.0_thesis_measurements_20261102.zip} (299\,841\,857 bytes, MD5 \texttt{177742116d2fc0a329f199cff3a2de19}); its tracked payload is the git archive of commit \texttt{cd66fd3} of the branch \texttt{release/v2.0\_thesis}, whose full hash is stored as the zip comment, and its split parts are committed as \texttt{922cddd}. Earlier revisions 5 and 6 of the same file name were superseded by revision 7, so the MD5 sum and the zip comment identify the file. Apart from \path{PACKING_NOTE.txt}, revision 7 adds to revision 4 (\path{openmx40gpu_v2.0_thesis_measurements_20260908.zip}, commit \texttt{40753d6}) only the cnc128 stand-alone GEMM measurements of 1 November 2026, the re-aggregated CSV files of Appendix~\ref{app:provenance}, and the Amdahl tables of the stand-alone tests recomputed from the full-precision means; these 28 added and changed files are also tracked individually in the repository (\path{work/gemm_micro/} and \path{work/reaggregate/} at commit \texttt{cd66fd3}) and are bundled, beside the split parts at commit \texttt{7af14b87}, as \path{cnc128_and_reaggregation_supplement_20261102.zip} (304\,540 bytes, MD5 \texttt{193c54ea927ab2255b8a5185413ae12e}). The archive contains the inputs, run scripts, individual outputs, manifests, environment records, aggregations, and provenance; the packing scope and the excluded items are described in \path{PACKING_NOTE.txt} and \path{EXTRAS_FILELIST.txt}. Large volumetric data, restart files, and executables are excluded, and the pseudopotential and basis data DFT\_DATA19 must be provided separately in the version specified by the inputs. The correspondence between the CSV files and analysis scripts of the existing performance figures and the aggregations of the present practical-convergence re-measurement is given in Appendix~\ref{app:provenance}.

\end{document}